%% file: alder-ICDSC-2022.tex
\documentclass[conference]{IEEEtran}

\input{header.tex}

\begin{document}

\title{\protoname: Unlocking blockchain performance by multiplexing consensus protocols\\
\thanks{Identify applicable funding agency here. If none, delete this.}
}

\author{
 \IEEEauthorblockN{
     Kadir Korkmaz\IEEEauthorrefmark{1}
     Joachim Bruneau-Queyreix\IEEEauthorrefmark{1}
     Sonia Ben Mokthar\IEEEauthorrefmark{2}
     Laurent Réveillère\IEEEauthorrefmark{1}
}

 \IEEEauthorblockA{\IEEEauthorrefmark{1} Univ. Bordeaux, CNRS, Bordeaux INP, LaBRI, UMR5800, F-33400 Talence, France}%
 \IEEEauthorblockA{\IEEEauthorrefmark{2} CNRS - LIRIS. Lyon, France}%

}


\maketitle

\input{section/abstract}


\input{section/introduction}
\input{section/related_work}

\input{section/goals_and_assumptions.tex}

\input{section/system_design}

\input{section/case_studies}

\input{section/evaluation}
\input{section/conclusion}

\bibliographystyle{IEEEtran}
\bibliography{IEEEabrv,zotero_autogenerated_bib,sample-base}

\end{document}

%% file: header.tex
\usepackage[T1]{fontenc} 
\usepackage[utf8]{inputenc}

\usepackage{cite}
\usepackage{amsmath,amssymb,amsfonts}
\usepackage{algorithmic}
\usepackage{graphicx}
\usepackage{textcomp}
\usepackage{xcolor}

\usepackage{xspace}
\usepackage{algorithmic}
\usepackage{tabularx,booktabs}
\usepackage[ruled,vlined]{algorithm2e}

\usepackage{amsthm}
\usepackage{siunitx} 

\usepackage[ddmmyyyy]{datetime}
\usepackage{lastpage}
\usepackage{fancyhdr}
\usepackage{balance}
\usepackage[abbreviations]{foreign}
\usepackage{pifont}
\usepackage{xspace}
\usepackage{booktabs}
\usepackage{makecell}
\usepackage{tabularx}
\usepackage{array}
\usepackage{soul}
\usepackage{subcaption}

\newcolumntype{P}[1]{>{\centering\arraybackslash}p{#1}}
\newcolumntype{M}[1]{>{\centering\arraybackslash}m{#1}}
\newcolumntype{C}[1]{>{\centering\arraybackslash}m{#1}}

\newcolumntype{Y}{>{\centering\arraybackslash}X}

\newcommand{\protoname}{\textsc{ALDER}\xspace}
\newcommand{\ba}{\textit{BA$\star$}\xspace}
\newcommand{\bba}{\textit{Binary}\ba}

\newcommand{\algorand}{Algorand\xspace}
\newcommand{\algorandplus}{Algorand++\xspace}
\newcommand{\rapidchain}{Rapidchain\xspace}
\newcommand{\rapidchainplus}{Rapidchain++\xspace}
\newcommand{\bitcoin}{Bitcoin\xspace}
\newcommand{\bitcoinplus}{Bitcoin++\xspace}

\newcommand{\bcp}{\textit{BCP}\xspace}
\newcommand{\bcplus}{\textit{BCP++}\xspace}

\IEEEoverridecommandlockouts

\def\BibTeX{{\rm B\kern-.05em{\sc i\kern-.025em b}\kern-.08em
    T\kern-.1667em\lower.7ex\hbox{E}\kern-.125emX}}

\newcommand\dingnum[1]{\ding{\numexpr181+#1}}

%% file: section/abstract.tex
\begin{abstract}

  
  Most of today's online services (\eg social networks, search engines, market places) are centralized, which is recognized as unsatisfactory by a majority of users for various reasons (\eg centralized governance, censorship, loss of control over personal data). Blockchain technologies promise a new Web revolution (Web 3.0) through the decentralization of online services. 
  However, one of the key limitations for this revolution to happen at a planetary scale is the poor performance of today's blockchains. 
  We propose in this paper \protoname, a solution for unlocking the performance of off-the-shelf leader-based blockchains by multiplexing their consensus protocol. 
  Our solution leverages the existence of multiple potential leaders to alleviate the bottleneck that exists at different levels of consensus protocols.
  To illustrate the benefits it brings to Blockchain performance, we apply \protoname to three representative blockchains, namely Algorand (Proof-of-Stake), RapidChain (Sharding-based) and Bitcoin (Proof-of-Work). 
  Our evaluation, involving up to 10,000 nodes deployed on 100 physical machines, shows that using \protoname can provide up to a 300\% improvement in both throughput and latency reduction.
  

\end{abstract}

%% file: section/introduction.tex
\section{Introduction}


%


Blockchains are decentralized, globally distributed, strongly consistent replicated systems that run across networks of mutually untrusted nodes. 
Since the emergence of the decentralized Bitcoin protocol~\cite{nakamotoBitcoin2008}, permissionless blockchains have demonstrated their ability to run arbitrary distributed applications~\cite{androulakiHyperledger2018} with the promise of supporting entire decentralized economies~\cite{giladAlgorandScalingByzantine2017}, business ecosystems across industries~\cite{noauthor_hyperledger_nodate}, decentralized service infrastructures~\cite{uriarteBlockchainBased2018} and have even the ambition of decentralizing the entire Web~\cite{raval2016decentralized}. Achieving the latter use cases requires efficient and performant blockchain solutions.
From the oldest blockchains (\eg Bitcoin) to the newest (\eg OHIE, Red Belly), the race towards more efficient and performant blockchains is still ongoing. While some blockchains aim at replacing the others, it is commonly admitted that there is no one blockchain that will fit the needs of the wide variety of distributed applications~\cite{belchior2022survey}. In this context, solutions that aim at improving the performance of off-the-shelf blockchains take a real sense, which is our aim in this paper. 

A key building block that is often considered a bottleneck in blockchain performance is the consensus protocol used to agree on the next block to append~\cite{guerraoui2019consensus}. 
There are a variety of these blockchain consensus protocols (\eg Proof-of-Work~\cite{nakamotoBitcoin2008,wood2014ethereum,eyalBitcoinNG2016,kogias2016enhancing,zamaniRapidChain2018,yuOHIE2020,abraham2016solida}, Proof-of-Stake~\cite{giladAlgorandScalingByzantine2017, davidOuroboros2018}, committee-based protocols~\cite{giladAlgorandScalingByzantine2017,abraham2016solida,kogias2016enhancing,zamaniRapidChain2018}).

In this paper, we focus on leader-based consensus protocols. 
These protocols, deployed at the heart of major blockchains (\eg Bitcoin, Ethereum, Hyperledger, Algorand), elect a leader that is responsible for proposing a block. 
In the process of electing a single leader, consensus protocols typically identify many potential leaders. 
We argue that blockchain protocols could directly benefit from the contribution of these candidate leaders to improve their performance by distributing the load and sharing the use of resources between them. 
More specifically, we claim that the contribution of multiple leaders to the addition of a block to the blockchain can either reduce confirmation latency or increase the amount of data added to a given block.

\noindent
\textbf{Contributions.}
We present \protoname a solution that empowers existing blockchains by multiplexing their consensus protocols, allowing multiple leaders to operate in parallel. 
Leaders in \protoname are elected by leveraging candidate leaders who are not exploited in the original consensus protocol. 
Then, the elected parallel leaders independently propose concurrent blocks containing disjoint sets of transactions, the union of which constitutes a macroblock to append. 
Since \protoname builds on existing blockchain protocols, the resulting protocols inherit the safety and liveness properties of the parent protocol.
To assess the effectiveness of \protoname, we apply its principles to three major blockchains: RapidChain~\cite{zamaniRapidChain2018}, a fast sharding-based blockchain, Algorand\cite{giladAlgorandScalingByzantine2017}, a scalable proof-of-stake blockchain, and Bitcoin, a well-known proof-of-work blockchain.
Our evaluation, involving up to 10,000 nodes deployed on 100 physical machines, shows that using \protoname can provide up to a 300\% improvement in both throughput and latency reduction.

\noindent
\textbf{Outline.}
We discuss related work in Section~\ref{sec:rw}.
We detail our system model and assumptions in Section~\ref{sec:goals_and_assumptions}.
We present \protoname and its building blocks in Section~\ref{sec:design}. 
We describe the use of \protoname on three representative blockchains in Section~\ref{sec:application}.
We evaluate the resulting systems in Section~\ref{sec:evaluation}, and conclude in Section~\ref{sec:conclusion}.

%% file: section/related_work.tex
\section{Related Work}
\label{sec:rw}


The past decade acknowledges numerous solutions aiming to improve the performance of blockchain consensus algorithms. For instance, BitcoinNG~\cite{eyal2016bitcoin} improves the throughput of the Bitcoin protocol by decoupling leader election from transaction serialization.
Similarly to Bitcoin, in BitcoinNG a leader is elected using cryptographic puzzles. The difference resides in the fact that the latter submits blocks until a new leader is elected. In this protocol, forks frequently happen during the switch between one leader to another. ByzCoin~\cite{kogias2016enhancing} also decouples leader election from transaction serialization as in BitcoinNG. In ByzCoin, the last $n$ puzzle solvers constitute a committee to decide on blocks. Committee members use communication trees to improve the communication complexity, and they use CoSi~\cite{syta2016keeping} to have a fast consensus. Algorand relies on cryptographic sortitions and verifiable random functions (VRFs) to sample a subset of nodes that will act as leaders and committee members. RapidChain~\cite{zamani2018rapidchain} partitions nodes into committees called shards that sustain disjoint blockchains. To boost performance, RapidChain employs an efficient synchronous consensus algorithm that decreases the cost of communication inside a committee. Similarly to RapidChain, Monoxide~\cite{wang2019monoxide}, Elastico~\cite{elasticoluu2016} and Omniledger~\cite{kokoris2018omniledger} also employ sharding to improve the transaction processing capacity of blockchains. 

Although all these propositions improve over the performance of existing blockchains, they remain specific solutions that can hardly be exploited for future blockchains. 
Instead, we present in \protoname a set of principles that can be applied to a variety of off-the-shelf blockchains to improve their performance.
To reach this objective, \protoname leverages primitives that have been successfully used in various consensus protocols. Specifically, we have been inspired by the following protocols when developing \protoname: Paxos ~\cite{lamport2001paxos} is one of the first crash fault-tolerant consensus algorithms, and Mencius\cite{barcelona2008mencius} improves the WAN performance of Paxos protocol by dividing the available sequence numbers between replicas. Hence, each replica leads a separate consensus instance of its own. 
BFT-Mencius~\cite{milosevic2013bounded} applies the approach of Mencius to the famous PBFT~\cite{castro1999practical} protocol.
Unlike Mencius, BFT-Mencius targets the bounded delay in such a way that each client request can be processed in a bounded time interval despite the presence of Byzantine nodes in the system. 
Mir-BFT\cite{stathakopoulou2019mir} is a recent solution that employs a similar technique to improve the WAN performance of the PBFT protocol. 
Unlike Mencius and BFT-Mencius, Mir-BFT employs a novel request bucket assignment technique to handle censorship attacks. In Mir-BFT, client requests are partitioned into a fixed number of buckets. Then, each bucket is assigned to one of the replicas. Assigned buckets are frequently rotated so that a replica cannot censor requests from specific clients. 
\protoname's principles are also inspired by the bucket mechanism to partition the space of requests between leaders as further discussed in Section~\ref{sec:design}.

OHIE\cite{yu2020ohie} is a recent blockchain protocol. OHIE uses cryptographic puzzles to elect leaders, and its Blockchain consists of up to 1000 parallel chains. In OHIE, nodes replicate all parallel chains. Like \protoname, nodes in OHIE propose small blocks that disseminate fast, and OHIE elects many leaders by decreasing the difficulty of cryptographic puzzles. In OHIE, an elected leader is assigned to one of the parallel chains in a publicly verifiable manner to append a block. Unlike \protoname, OHIE does not partition the hash space of transactions to handle redundant transaction processing; therefore it might suffer from appending the same transaction several times. 

Red Belly\cite{crain2021red} is another recent blockchain proposition that relies on a novel Democratic Byzantine Consensus (DBFT) algorithm. Red Belly considers a permissioned system in which only subset of nodes is allowed to be proposer; therefore, proposers are not elected. As \protoname, DBFT has proposers that propose disjoint set of transactions to append to the ledger. All proposed disjoint transactions can be appended at the end of consensus as a super-block. Red Belly employs a partial validation mechanism to share the cost of transaction validation as nodes only validate a subset of transactions.

%% file: section/goals_and_assumptions.tex
\section{System model and objectives}
\label{sec:goals_and_assumptions}

The work presented in this paper aims to improve the performance of existing leader-based blockchain consensus protocols by optimizing resource usage on the critical path to consensus decisions.
Our approach does not consist of inventing new consensus protocols. 
Conversely, we observe that blockchains can suffer from bottlenecks at various levels, from the efficiency of block dissemination techniques, to the average waiting time between the resolution of two crypto-puzzles, to the communication complexity of committee-based consensus.
\protoname proposes to alleviate these bottlenecks by multiplexing the execution of the consensus protocol, involving more leaders in the different steps leading to consensus decisions.
These additional leaders are identified from the set of candidate leaders in the original election mechanism.
In this work, we consider a subset of blockchain protocol models that incorporate leader election, block dissemination and consensus rules operating under partial synchrony.
Our work excludes leaderless consensus algorithms~\cite{antoniadis2021leaderless} and leaderless blockchain models such as Avalanche~\cite{rocket2019scalable}. 

\protoname consists of a set of reusable principles that, when applied to the different components of an existing blockchain protocol, improve its performance in terms of throughput or transaction latency.
The resulting protocol inherits the safety and liveness properties of the original consensus protocol by holding on to the messaging, data structures, and decision making rules of the underlying protocol.

%% file: section/system_design.tex
\section{\protoname: System Design}
\label{sec:design}

\begin{figure}[t]
    \centering
    \includegraphics[width=.95\columnwidth]{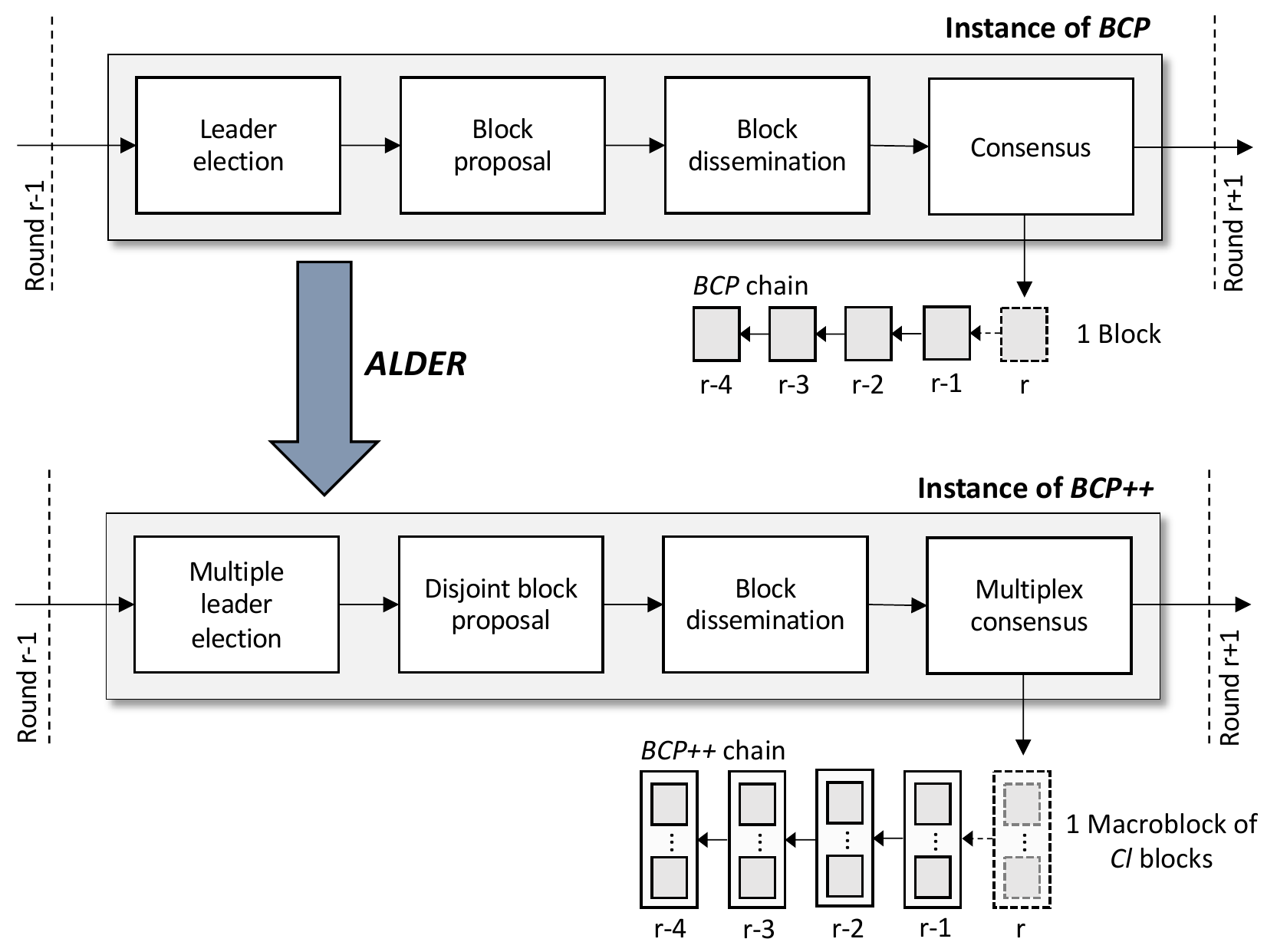}
    \caption{Multiplexing a blockchain \bcp using \protoname}
    \label{fig:general_case}
    \vspace{-4mm}
\end{figure}

In this section, we present the set of \protoname principles that allow to multiplex a leader-based Blockchain Consensus Protocol (\bcp) and increase its performance. 
The resulting blockchain consensus protocol is referred to as \bcplus in the remainder. 
Figure~\ref{fig:general_case} illustrates this transformation on a \bcp in its canonical form with: leader election, block proposal, block dissemination and consensus. 
In the blockchain models considered, the leader election mechanism identifies a node (sometimes several, either by design principle or by side effect) that must propose a candidate block to the rest of the nodes in the system.
Once disseminated to the other nodes, an agreement must be reached to consider the proposed block as a valid extension of the \bcp blockchain.

\newcounter{c}

\protoname extends \bcp to \bcplus using the following four principles:
(1) In each round, \bcplus elects multiple leaders by extending the election mechanism of the original \bcp, 
(2) leaders propose blocks that contain disjoint sets of transactions by 
(3) partitioning a transaction hash space into buckets and assigning each leader to one of the buckets in an unforgeable and publicly verifiable way.
Finally, (4) \bcplus runs a multiplexed version of the \bcp consensus to produce a decision on a composition of the proposed blocks.
We call this last composition \textit{macroblock}.
The nodes in the system wait for this decision to aggregate the agreed-upon list of blocks and actually add a \textit{macroblock} to the chain before moving on to the next protocol round.
We now detail the principles of \protoname and how they articulate together.


\subsection{Multiplexed consensus and macroblocks}
A macroblock is a logical composition of up to $Cl$ blocks of size $bs$ bytes, with $Cl$ a bootstrap parameter that defines the \textit{concurrency level} of \bcplus. The resulting macroblock has a size of $Cl \times bs$ bytes.
Each block in the macroblock contains the hash of the previously appended macroblock.

The consensus of \bcplus produces a decision on the composition of the next macroblock that should be appended to the chain.
This decision consists of a vector of up to $Cl$ hashes of candidate blocks.
Each node in the system listens for the blocks proposed by the set of leaders, and locally builds a macroblock based on the consensus decision produced.

\begin{figure}[ht]
\vspace{-2mm}
  \centering
  \includegraphics[width=.95\columnwidth]{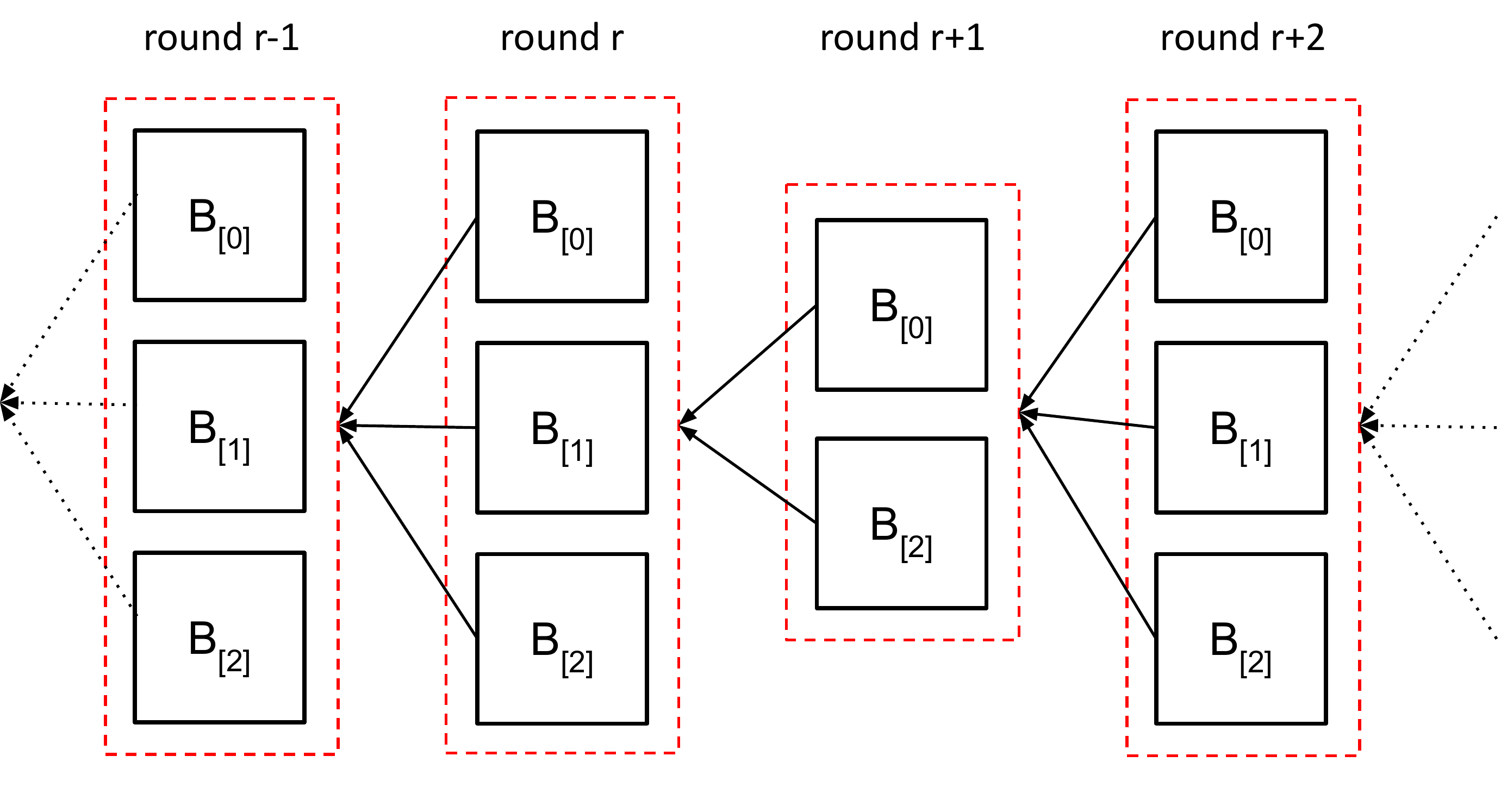}
  \caption{\bcplus macroblockchain structure}
  \label{fig:alder_blockchain_structure}
  \vspace{-3mm}
\end{figure}

Figure~\ref{fig:alder_blockchain_structure} depicts an example of a chain of macroblocks with a concurrency level $Cl$ of 3.
Dashed lines represent macroblocks, and solid squares indicate the composing blocks.
\protoname does not require that each macroblock contain exactly $Cl$ blocks as shown in the figure.
Indeed, the blockchain protocol model considered for \bcp includes existing protocols that can produce empty blocks, such as Algorand~\cite{giladAlgorandScalingByzantine2017}.
In addition, consensus on some of the proposed blocks may not be reached.

\subsection{Transaction duplication and duplication attacks}

We leverage the leader election mechanism of \bcp to involve more leaders in the blockchain consensus protocol and increase its performance.
More specifically, all elected leaders propose candidate blocks to form the next macroblock. 

However, appending multiple blocks of transactions in a given round poses the problem of duplicate transactions and duplicate attacks.
Indeed, in a simple approach, leaders could create blocks with any transaction they received.
As a result, some blocks could include transactions that would also appear in some of the blocks proposed by other leaders.
This would reduce the throughput gain envisioned in our approach and increase the complexity of the transaction execution phase.

Duplication of transaction  also opens the doors for duplication attacks by an adversary controlling Byzantine nodes.
When some of these nodes get elected, the adversary may wait to learn about blocks proposed by the honest leaders and have its Byzantine leaders proposing blocks containing the same transactions, thus reducing the performance gain of \bcplus.



To cope with this problem, \protoname incorporates the principles of partitioning the transaction hash space and proposing disjoint blocks.



\subsection{Transaction hash space partitioning and bucket assignment}

\begin{figure}
  \centering
  \includegraphics[width=4cm,keepaspectratio]{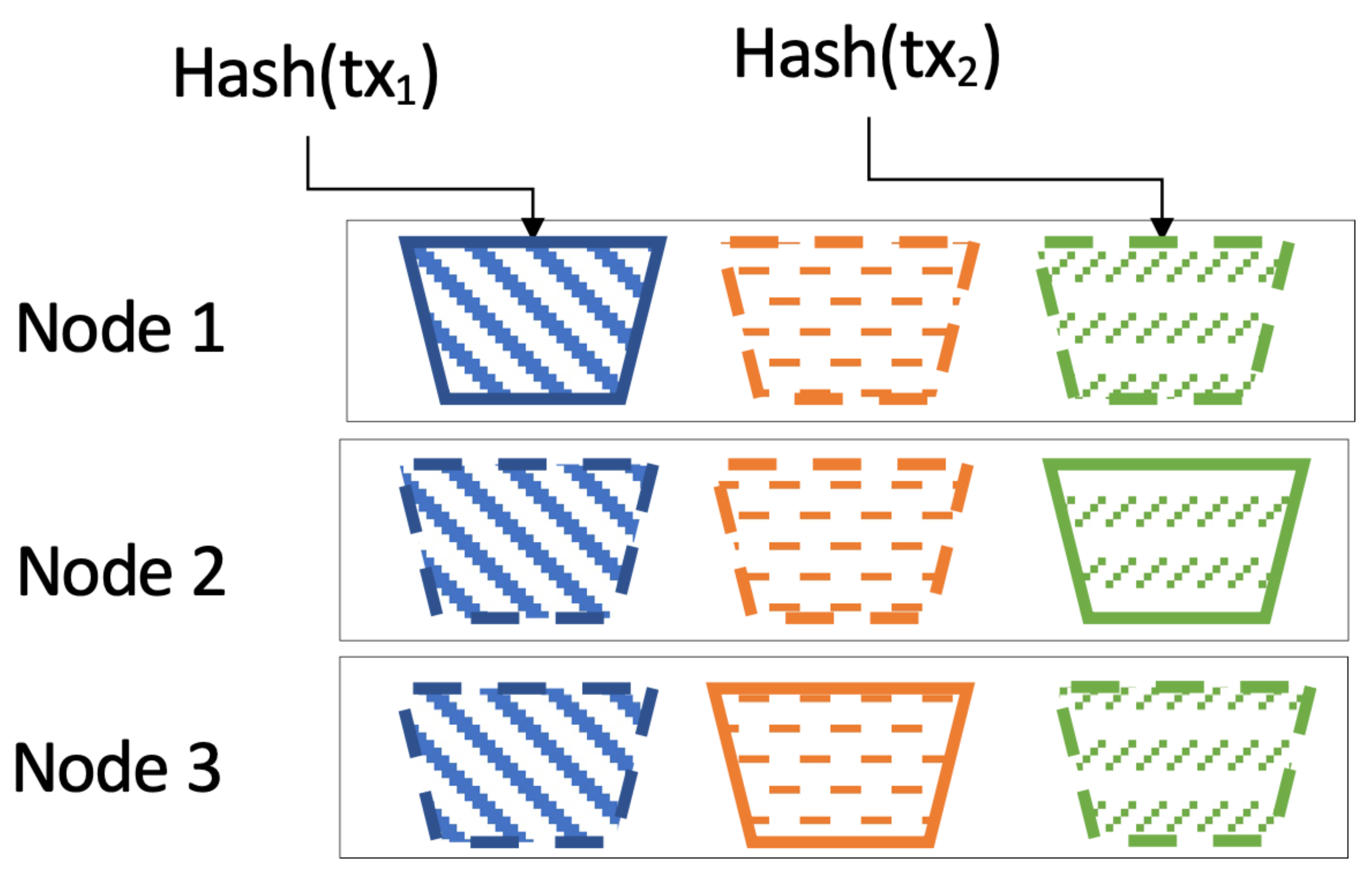}
  \caption{Mapping transactions to a bucket with $Cl=3$. Solid lines represent the assigned bucket for a given node. Hash(tx$_1$) is mapped to the first bucket, assigned to node 1. Hash(tx$_2$) is mapped to the third bucket, assigned to node 2.}
  \label{fig:bucket_assignment}
  \vspace{-2mm}
\end{figure}

Nodes partition the transaction hash space into $Cl$ non-intersecting buckets of equal size. 
Each leader is assigned a bucket of transactions in an unforgeable and publicly verifiable way during the leader election phase.
To avoid duplication attacks, leaders cannot decide on which bucket they are assigned to.
Each block proposed by a leader must contain exclusively transactions whose hashes fall within the bucket assigned to the leader.
In this way, any node can validate the correct fabrication of a received block by checking its contents against the bucket assigned by the block proposer.
Figure~\ref{fig:bucket_assignment} illustrates the mapping of transactions to assigned buckets, with $Cl=3$.


%% file: section/case_studies.tex
\section{Case studies: application on three blockchain protocols}
\label{sec:application}

In this section, we show the application of \protoname's principles on three different permissionless blockchain consensus protocols: \algorand, \bitcoin and \rapidchain.
Each of these blockchains is representative of a different blockchain family. 
\bitcoin is the original PoW-based blockchain protocol described by Nakamoto~\cite{nakamotoBitcoin2008}. 
\rapidchain illustrates one of the most throughput-efficient approaches for sharded blockchain protocols.
Finally, \algorand is among the most scalable stake- and committee-based blockchain protocols achieving minimal transaction confirmation latency.





\input{casestudies/algorand.tex}
\input{casestudies/rapidchain.tex}

\input{casestudies/bitcoin.tex}

%% file: casestudies/algorand.tex
\subsection{\algorandplus: Applying \protoname to \algorand}

\algorand~\cite{giladAlgorandScalingByzantine2017} is among the most scalable PoS-based permissionless blockchain.
To scale the consensus to many users, it relies on a cryptographic sortition mechanism that randomly and in non-interactive way selects committees of nodes.
Committees are charged either to propose blocks or to contribute to a Byzantine agreement protocol called \ba to reach a consensus on one of the proposed blocks.
Algorand has a low transaction confirmation latency of the order of a minute. 
Despite the impressive performance of Algorand (125-fold throughput improvement compared with the Bitcoin), this protocol still suffers from performance limitations. 
In particular, its performance drops dramatically with large block sizes.
Indeed, the time of gossiping blocks in the network largely dominates the duration of \ba. 
This long gossip is an important limitation to increase throughput. 
Increasing the block size would only increase the confirmation latency and keep the throughput at its highest level in the best case.


To make a better use of the network as well as the proposed blocks, we consider \protoname's principle as a sound approach. 

\subsubsection{\algorand protocol}
\begin{figure}
    \centering
    \includegraphics[width=\columnwidth]{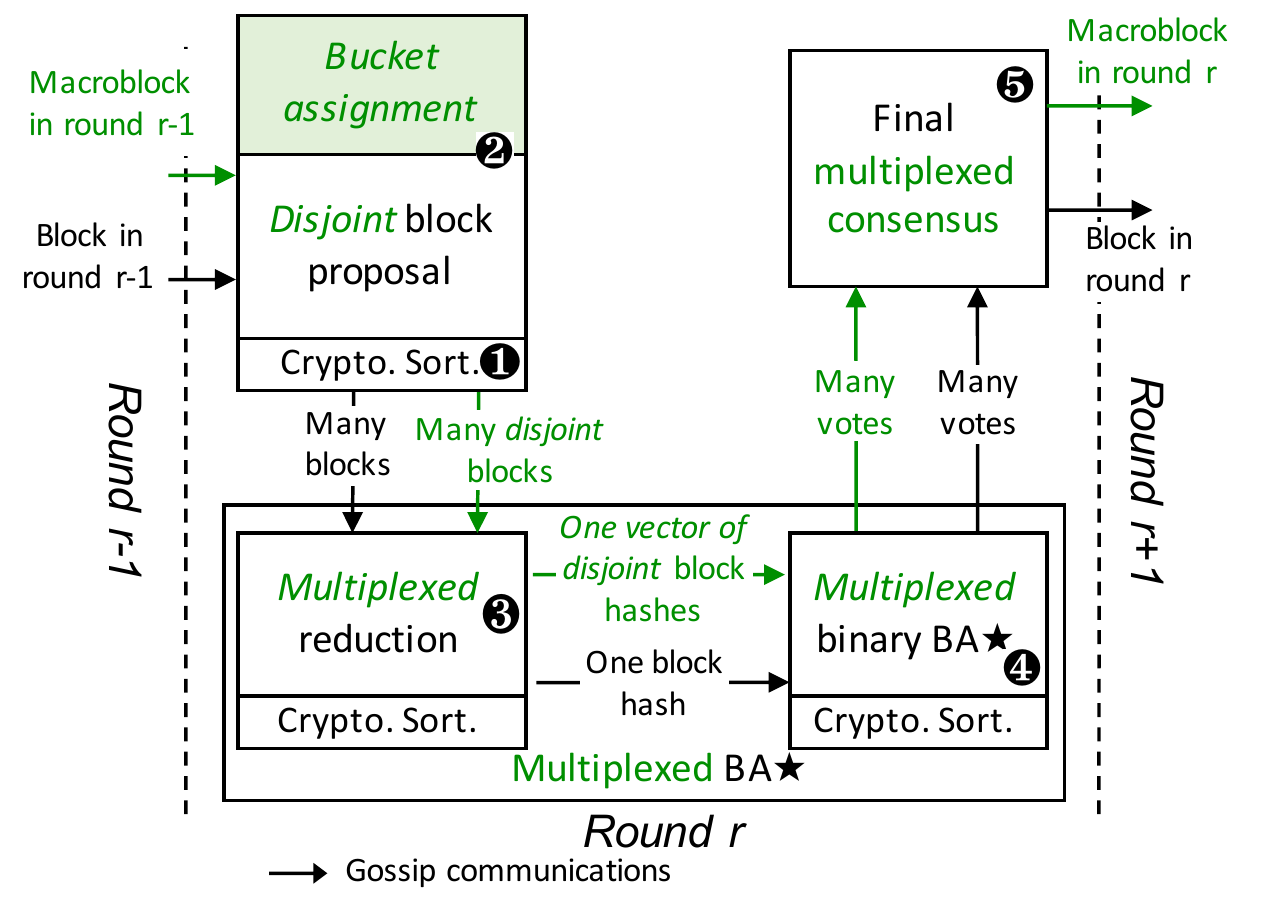}
    \caption{\algorand and \algorandplus round structures. Black texts, arrows and boxes depict the round structure of \algorand while green elements illustrate \algorandplus}
    \label{fig:algoplusround}
    \vspace{-2mm}
\end{figure}

Figure~\ref{fig:algoplusround} depicts the five main steps of an \algorand round.
First (step \dingnum{1}), each node executes a cryptographic sortition to determine whether it belongs to some of the committees responsible for conducting the three following steps of the protocol: the block proposal step, the reduction and binary agreement steps of \algorand's Byzantine agreement \ba.
In the second step (step \dingnum{2}), a first committee of nodes, \ie leaders, proposes blocks to be appended to the blockchain. 
Then (step \dingnum{3}), \algorand's Byzantine agreement procedure \ba reduces the problem of agreeing on one of many blocks hashes to agreeing on either a selected block hash or a default empty block hash, and reaches consensus via a binary agreement called \bba (step \dingnum{4})
Finally (step \dingnum{5}), every node counts votes cast during the \ba phase to learn about the outcome of the agreement procedure a final consensus.
To prevent an adversary from learning the identity of committee nodes and forging targeted attacks, nodes determine if they belong to specific committees through an independent and non-interactive cryptographic sortition procedure.
In addition, committees are different for each step of the protocol to prevent targeted attacks on committee members once they send a message.
Regarding communications, \algorand relies on gossip.
Each node selects a small random set of peers to which it transmits block and protocol messages. 

\ul{\textit{Cryptographic sortition (leader election):}}
Cryptographic sortition enables an expected fraction of nodes to be selected at random based on their weights (\ie, their currency stake in the system) for a specific committee of the protocol in a publicly verifiable, and non-interactive manner.
Each node runs the cryptographic sortition procedure for each of the committees by computing a has value from a Verifiable Random Function~\cite{micaliVerifiableRandomFunctions1999} (VRF)
\footnote{
    $VRF$ is a public-key pseudorandom function that provides proofs that its outputs were calculated correctly. 
    The owner of the secret key $sk$ can compute the $\langle value,\pi \rangle \leftarrow VRF_{sk}(x)$
    on any input $x$ to produce a hash value function value and an associated proof $\pi$.
    Any node in possession of the associated public key can $\pi$ check that this value was indeed calculated correctly with the input $x$, without using the secret key.
    }
    of its private key and seed obtain from the last block of the blockchain.
If the function's output indicates that the node is chosen, it returns a short \textit{proof} string that proves this node’s committee membership to other nodes, which the node can include in its messages.

\ul{\textit{Block proposal:}}
All nodes execute the cryptographic sortition to determine whether they are selected to propose a block in a given round.
The sortition is designed to elect an expected number of $\tau_{proposer}$ block proposers.
At a high level, the sortition for the block proposal step ensures that an expected $\tau_{proposer}$ number of nodes are selected at random, weighted by their account balance. 
It also provides each selected node with a priority, which can be compared across users, and a proof of the selected node's priority. 
Since the selection is random, there may be multiple nodes selected to propose a block, and the priority determines which block every node should adopt. 
The selected nodes create and distribute their block of pending transactions through the gossip protocol, along with their priority and proof. 
To reduce unnecessary communications, and because only one of the proposed blocks will be appended to the chain, each node prioritizes blocks based on the priority of the block proposer.
Nodes wait a certain amount of time to receive priority messages and blocks (respectively 10 seconds and 1 minutes as empirically determined by the authors~\cite{giladAlgorandScalingByzantine2017}).
If a node does not receive a block within this delay, it proceeds to the protocol step considering an empty block.





In each round, block proposers add to their blocks a seed proposal required for the sortition process in the next round. 
The proposal seed is determined as the verifiable random hash value of the last round seed concatenated with current round index $r$, \ie $\langle seed_r,\pi \rangle \leftarrow VRF_{sk}(seed_{r-1} || r)$.

\ul{\textit{Consensus:}}
Algorand relies on a Byzantine agreement procedure \ba to reach consensus on a single block.
This procedure is composed of two phases during which each committee member contributes by casting a vote on block hash value, and every node in the system counts vote results.

The first phase is a reduction, which transforms the problem of agreeing on one block among many to agreeing on either an empty block hash or a proposed block hash.
The reduction phase consists of precisely two steps and requires the contribution of 2 committees.
Members of the first committee vote for the block hash with the highest priority they have received.
If a block hash receives a majority of votes in the first step, then members of the second committee will vote for this block hash; otherwise, they vote for the hash of a default empty block.
The reduction phase ensures that at most one non-empty blocks will be selected to be sent to \bba.

The second phase consists in executing a binary Byzantine agreement procedure \bba on the hash value passed on to \bba by the reduction phase and an empty block hash value.
This phase takes a variable (but bounded) number of steps to complete depending on the network synchrony and the honesty of the highest-priority proposer. 
Nodes continue to count votes as in the reduction phase, and the committee members of each step cast votes on block hashes.
In each step of \bba, when a node receives more than $T \times \tau$ votes for some value, it votes for that same value in the next step.
If a block hash receives a super majority of votes, a final consensus is reached and nodes can move on the next round. 
On the other hand, if no hash value receives a super majority,

\subsubsection{\algorandplus protocol}

We now apply \protoname's principles on \algorand to create \algorandplus in which multiple leaders can independently propose concurrent blocks containing disjoint sets of transactions, and grow the chain by appending a subset of the proposed blocks in every round. 
Figure~\ref{fig:algoplusround} depicts in green color the global course of a \algorandplus round. 
First (step \dingnum{1}), each node executes the cryptographic sortition to determine whether it is elected as a leader to contribute to the block proposal step as well as the other steps of \ba.
When elected, each leader also learns which bucket of transactions it should use to build a block from. 
Doing so, the leaders can submit propose messages and blocks containing \textit{disjoint} sets of transactions (step \dingnum{2}) along with a reference to the previous macroblock.
In the reduction step is \textit{multiplexed} so that committee members will reduce the problem of agreeing on a set of blocks to agreeing either on a \textit{vector} of disjoint block hashes or on the hash of an empty block (step \dingnum{3}) in precisely two steps. 
Finally, the multiplexed binary Byzantine agreement (step \dingnum{4}) reaches consensus one of the inputs provided by the reduction phase.
Once consensus is reached (step \dingnum{5}), nodes gather the blocks \algorandplus has agreed on and build the \textit{macroblock} corresponding to the consensus decision before appending it to the chain.

\ul{\textit{Bucket assignment and multiple leader election:}}
Electing multiple leaders is already part of the original protocol. 
Assigning buckets in an unforgeable and publicly verifiable way is performed by leveraging the original cryptographic sortition mechanism. 
More specifically, the assigned bucket index is computed from the VRF's hash value produced during the cryptographic sortition procedure modulo the concurrency level $Cl$.
Because \algorandplus decides on a vector of blocks and not one block only, the number of block proposers, \ie leaders, $\tau_{proposer}$ should be sufficiently large to avoid unassigned buckets.
A too low $\tau_{proposer}$ value could lead to buckets being frequently unassigned which would not only hinder the envisioned throughput gain, but would also lead to transactions with specific hash being unfairly ignored for some period.
To devise an appropriate value for $\tau_{proposer}$, we rely on the uniform distribution of bucket assignment deriving from the use of hash functions in the cryptographic sortition.
In other words, each node has an equal chance of being assigned one bucket over another; a bucket is assigned to at least one proposer with the following probability $1 - (1-\frac{1}{Cl})^{\tau_{proposer}}$; and all buckets are assigned to at least one proposer with probability $\sum_{i=0}^{Cl} (-1)^{Cl-i}\binom{Cl}{i} \big( \frac{i}{Cl} \big)^{\tau_{proposer}}$.

\ul{\textit{Disjoint block proposal: }}
The block proposal step of \algorandplus is very similar to that of \algorand, except that nodes can be assigned the same transaction bucket as other nodes.
Because only one proposed block per bucket index will be appended to the chain, \algorandplus extends the original protocol to reduce unnecessary communications.
Similarly to \algorand, each block proposer sends two kinds of messages to help other nodes decide on which block can be safely discarded for each specific bucket.
The first message is the proof of selection and a priority hash value used, this time, to compare the proposed blocks composed of transactions originating from the same bucket index. 
The priority hash is derived from hashing the VRF hash output concatenated with a publicly verifiable information of the node's stake in the system, and the assigned bucket index.
The second message is the block itself.

The seed generation mechanism is also different.
Indeed, in \algorand, block proposers add to their blocks a seed proposal required for the sortition process in the next round. 
This proposed seed is determined as the verifiable random hash value of the last round seed concatenated with current round index $r$, \ie $\langle seed_r,\pi \rangle \leftarrow VRF_{sk}(seed_{r-1} || r)$.
In \algorandplus, each elected node proposes a partial seed by executing the VRF function on the concatenation of the seeds from all blocks in the previous macroblocks. 
That is $\langle seed_r,\pi \rangle \leftarrow VFR_{sk}(seed_{r-1,1}||...||seed_{r-1,Cl}||r)$. 
Doing so, the next round seed is only revealed when the consensus \ba decides on the composition of the next macroblock.
Then, the actual seed used to run crytographic sortition is the hash of the seeds from all the blocks composing the last macroblock.

\ul{\textit{Multiplexed consensus and macroblocks:}}
The multiplexed \ba agreement protocol takes as input a list of block hashes and produces a vector of block hashes that compose the macroblock to be appended at the end of a round.
As the final decision could be a vector containing hashes of empty blocks, \algorandplus grows a chain of macroblocks of possibly different sizes.
Based on this decision, each node locally builds the agreed-upon macroblock by concatenating the associated blocks.
Block proposers in the next round will compute the hash of this macroblock to continue the chain. 

%% file: casestudies/rapidchain.tex
\subsection{\rapidchainplus: Applying \protoname to \rapidchain}



\rapidchain is among the most efficient sharded permissionless blockchain protocol.
\rapidchain splits the system in $k$ non-intersecting committees of $m$ nodes.
Each committee is in charge of maintaining and growing a specific shard of the blockchain.
\rapidchain employs an optimal intra-committee synchronous consensus algorithm to achieve very high throughput via a novel gossiping protocol for large blocks, and a provably-secure reconfiguration mechanism to ensure robustness. 
Using an efficient cross-shard transaction verification technique, \rapidchain avoids gossiping transactions to the entire network.


\begin{figure}
    \centering
    \includegraphics[width=\columnwidth]{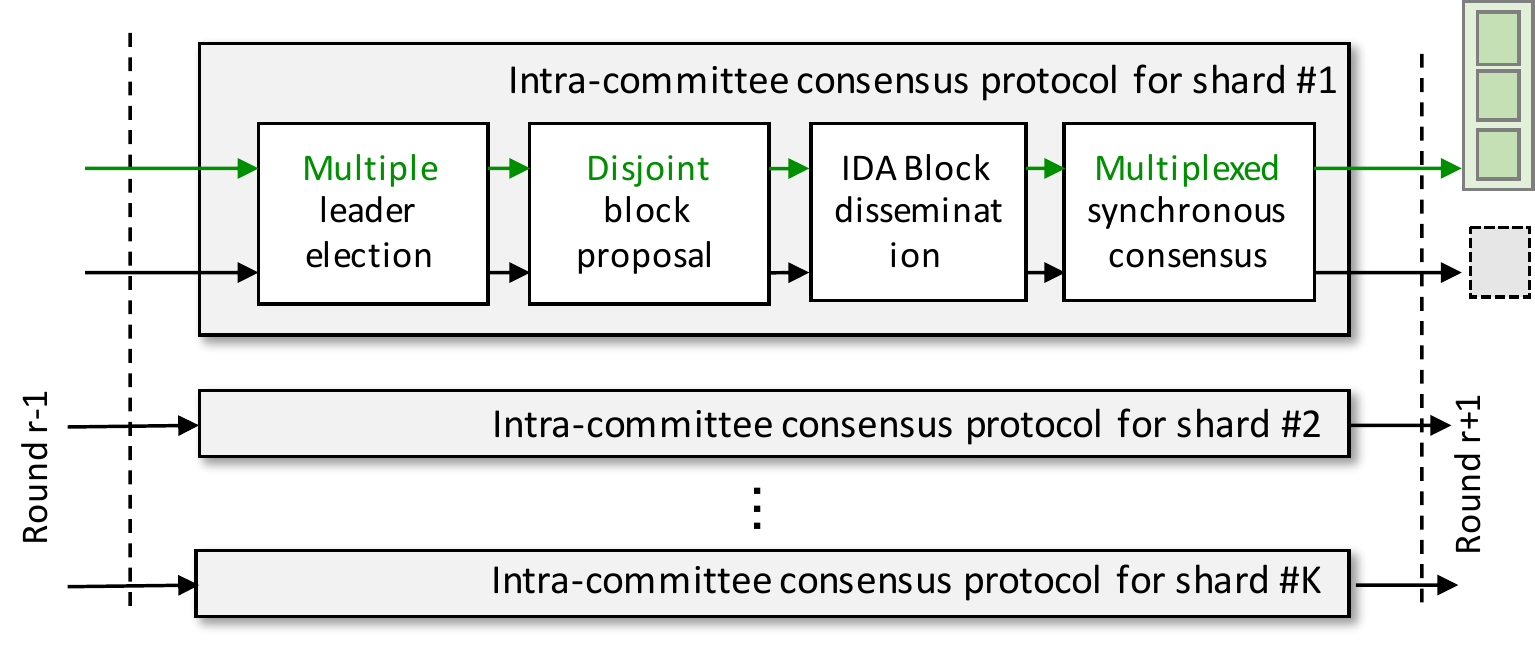}
    \caption{\rapidchain and \rapidchainplus round structures. Black texts, arrows and boxes depict the round structure of \rapidchain while green elements illustrate \rapidchainplus}
    \label{fig:rapidplusround}
    \vspace{-2mm}
\end{figure}




\subsubsection{\rapidchain protocol}
\rapidchain proceeds with successive day-long epochs consisting of multiple rounds of consensus. 
At the end of each epoch, a reconfiguration phase is executed.
Figure~\ref{fig:rapidplusround} depicts the course of a round. 
First a leader is elected in each committee in a deterministic manner, then the leader builds a block and gossip it to the committee with a specific gossip protocol inspired from IDA.
Finally, the committee reaches consensus on the proposed block via an intra-committee synchronous algorithm.

\ul{\textit{Reconfiguration and epoch-based committees:}}
At the end of every epoch, a reconfiguration phase driven by a specific reference committee $C_R$ allows nodes to join or stay in the existing committees.
$C_R$ agrees on a reference block consisting of the list of all active nodes for that epoch as well as their assigned committees, and appends the latter block to each shard.
To do so, $C_R$ runs a distributed random generation protocol to produce an unbiased random value called epoch randomness. 
Once generated, the epoch randomness is used to define a cryptographic puzzle challenge that each node wishing to join or stay in the system should solve within 10 minutes.
$C_R$ validates the received cryptographic puzzle solutions, assigns each member to a sharding committee, and informs other committees by publishing a configuration blocks.



\ul{\textit{Inter-committee routing and cross-shard verification:}}
Committee members wait for external users to submit their transactions. 
To split transactions onto the different shards, \rapidchain deterministically assigns each transaction to a committee by hashing the transaction identifier to a committee number. 
Nodes receiving transactions forward them to their destination committee using a routing protocol inspired by Kademlia~\cite{maymounkov2002kademlia}.

Committee members batch several transactions into a block and run an intra-committee consensus in order to append it to their shard.
Before the block can be appended, the committee verifies the validity of every new transaction in the block by verifying that the associated input transactions actually record enough previously-unspent currency.
Since transactions are stored into disjoint shards held by different committees, committee members need to communicate with the corresponding input committees to ensure that the input transactions exist in their shards.

\ul{\textit{Leader election:}}
For each round and each committee, \rapidchain deterministically elect a single leader from the committee members by using the epoch randomness and round number.
The leader is responsible for driving the consensus protocol.
Before that, it gathers the transactions it has received in a block and gossips it using the IDA gossip protocol. 

\ul{\textit{Intra-committee consensus:}}
In order to append one block per shard, \rapidchain proposes an intra-committee consensus protocol based on the synchronous protocol of Ren et al.~\cite{ren2017practical}, which is resilient to $f=1/2$ of the committee size.
The leader initiates the consensus on the header of the block it has gossiped. 
The protocol executes the following four steps in which all communications are signed by the sender's private key to provide authentication and integrity of messages.
(1) The leader submits a \textit{propose} message containing the block header,
(2) all nodes receiving the propose message send an \textit{echo} message containing the same block header to the entire committee. 
(3) If an honest node sees another version of the block header in one of the \textit{echo} messages it received, then it knows that the leader is corrupt, and it gossips a \textit{pending} message containing an empty block hash.
4) Finally, if an honest node receives $mf+1$ echo messages of the same and only block header, it decides on this block header and submits an \textit{accept} message including the $mf+1$ received echo messages.

\subsubsection{\rapidchainplus protocol}
We apply \protoname's principle for the consensus protocol executed by each committee.
Figure~\ref{fig:rapidplusround} depicts the course of a \rapidchainplus round:
First, \rapidchainplus uses the deterministic leader election protocol of \rapidchain to elect $C_l$ leaders.
Then, each leader builds and gossips concurrent blocks containing disjoint sets of transactions. 
Finally, the intra-committee synchronous consensus protocol is executed to decide on a vector of block headers.



\ul{\textit{Bucket assignment and disjoint block proposal:}}
Elected leaders are assigned buckets deterministically in the order of election: the first leader is assigned to bucket 0, the last leader is assigned to bucket $Cl-1$. 
An elected leader uses the assigned transaction bucket to build and gossip a block with the IDA gossip protocol.
Because leader election results and bucket assignment are publicly verifiable, any node can verify the correctness of the proposed blocks.
The leader submits the block message, and starts synchronous consensus by submitting a propose message. A propose message contains a single block header.

\ul{\textit{Multiplexed intra-committee consensus:}}
\rapidchainplus extends the synchronous consensus algorithm of \rapidchain in order to decide on a vector of block headers. 
Specifically, \textit{echo} and \textit{accept} votes contains the latter vector. 
Each committee member awaits up to $Cl$ \textit{propose} votes until the end of the synchronous round timeout, and then submits \textit{echo} votes that contain for the received block headers.
If it receives more than one version of a block header from a specific leader and bucket, it submits a \textit{pending} message which contains an empty block header. 
When receiving $mf+1$ echo votes for the same vector of block headers, a committee member sends an \textit{accept} message for this vector.


%% file: casestudies/bitcoin.tex
\subsection{Bitcoin with \protoname: \bitcoinplus}
In this section, we present Bitcoin and \bitcoinplus protocols. The latter is the improved version of the Bitcoin protocol with \protoname. 

\subsubsection{\bitcoin protocol} \label{bitcoin_overview}
Bitcoin is a Proof of Work (PoW) based blockchain system that relies on hash puzzles (cryptographic puzzles) to elect leaders. 
The global course of a round in \bitcoin includes the following steps:
(1) each node builds a block of transactions containing in its header the hash of the previously appended block and a nonce value of its choice.
(2) Then each node tries to find a nonce value that satisfies the following cryptographic puzzle. The hash value resulting from the evaluation of a cryptographic hash function over the block header should be lower than a threshold, also known as difficulty. 
The difficulty is adjusted by the system so that a single solution to the cryptographic puzzle is found every 10 minutes on average.
(3) when a node finds a solution to the cryptographic puzzle, it is considered a leader in bitcoin and sends the block it has built via a gossip protocol to its neighbors. 
(4) Upon receiving a block, each node verifies the validity of the solution and appends the block to its blockchain.
(5) If two different solutions are found for the same cryptographic puzzle, there exists two valid candidate blocks competing to extend the blockchain. 
In this case, nodes continue trying to solve the next puzzle considering the first block they have received, potentially extending the chain on two different paths, usually referred to as forks.
\bitcoin's consensus rely on \textit{the longest chain rule} to resolve forks by considering the highest amount of the work done the nodes in solving the puzzles and appending blocks to the chain.
In Bitcoin, nodes submit blocks of 1 MB. The small block size and predetermined long block intervals (on average 10 minutes) defined by the cryptographic puzzle difficulty are the main bottlenecks that limit the throughput of the Bitcoin system.



\subsubsection{\bitcoinplus protocol}
\bitcoinplus allows multiple nodes to independently propose concurrent blocks containing disjoint sets of transactions and grow the chain by appending a subset of the proposed blocks in every round. 

\ul{\textit{Macroblocks:}}
Contrarily to the previous blockchains, macroblocks in \bitcoinplus are composed of exactly $Cl$ blocks.
Indeed, as the consensus termination of \bitcoin is probabilist, and nodes cannot be sure that a given block is final and will never be removed from the blockchain. 
For this reason, nodes in \bitcoinplus only consider that a round in terminated when they can build a macroblock fully with $Cl$ blocks.

\ul{\textit{Multiple leader election:}}
Similar to its parent, \bitcoinplus relies on cryptographic puzzles to elect multiple leaders. 
To elect multiple leaders within the same time interval (10 minutes) on average, \protoname requires to decrease the difficulty of the cryptographic puzzle.
More precisely, to obtain $Cl$ leaders at the same time interval as \bitcoin, the difficulty must be divided by $Cl$.


\ul{\textit{Bucket assignment and disjoint block proposal:}}
\bitcoinplus assigns transaction buckets nodes at the time they solve a cryptographic puzzle for a macroblock. 
Specifically, each node builds a macroblock of $Cl$ blocks. The macroblock includes a nonce value, the hash of the previous macroblock, and a commitment scheme (a Merkle Tree) to commit each block to the built macroblock.
Then, the node tries to find a nonce value that satisfies the following cryptographic puzzle. 
The hash value resulting from the evaluation of a cryptographic hash function over the macroblock header should be lower than a target difficulty.
When a solution is found, the node learns the bucket index it was assigned by computing the modulo operation of the hash value against the concurrency level $Cl$. It then gossips the block corresponding to designated bucket along with the macroblock header used in the cryptographic puzzle as a proof.


\ul{\textit{Adressing competing blocks}}
Because $Cl$ leaders are expected every 10 minutes on average and, because the bucket assignment process may assign one bucket to more than one leader, then the time required to obtain $Cl$ blocks for each bucket could be greater than 10 minutes.
This could inhibit the performance gain envisioned by \protoname.
\bitcoinplus addresses this issue by introducing a sibling mechanism that enables to collect blocks competing for the same bucket index.

When building its block, each node adds a \textit{siblings} field to each block header.
This field is a $Cl$-long byte array that contains the hashes of blocks proposed by other nodes in the same round.
When the node receives a block for the same bucket index of the block it produces (or of the block it received), it looks at the siblings field of each block.
The node then deterministically use the next empty slot in the siblings field to link the received block to the others it has.



\ul{\textit{Multiplexed longest chain consensus rule}}
The siblings field decreases the probability of loosing the contribution of two blocks competing for the same bucket index.
However, it does not eliminate it completely because the latter mechanism could increase the likelihood of fork occurence.
Indeed, different nodes might accept different blocks for the same bucket index in a round. 
To handle this case, \bitcoinplus extends the \textit{longest chain consensus rule} with a priority consensus rule.
When receiving several blocks for the same bucket index, blocks are prioritized by their hash values (smallest wins) to help the node decide which one to consider.

%% file: section/evaluation.tex
\section{Evaluation}
\label{sec:evaluation}

In this section, we evaluate the extent of \protoname's ability to improve performances of the three blockchains we considered.
We first present our implementation and evaluation environment (Section \ref{implementations_and_experiments}) before presenting the performance of \algorandplus, \rapidchainplus and \bitcoinplus in sections \ref{algopp_performance}, \ref{rapidpp_performance} and \ref{bitcoinpp_performance}, respectively.

\subsection{Implementation and evaluation environment} 
\label{implementations_and_experiments}

We implemented all baseline protocols and their multiplexed versions using Golang. 
The experiments presented in this section were carried out using the Grid'5000~\cite{bolze2006grid} testbed.  
We used powerful physical machines, each with 18 cores, 96~GB of memory, and a 25 Gbps intra network connectivity link. 
In all experiments, we emulate wide-area network conditions as in major blockchain propositions~\cite{giladAlgorandScalingByzantine2017,zamaniRapidChain2018}: 
we cap the bandwidth for each process to 20~Mbps, and we add a one-way latency of 50 milliseconds (Round-Trip Time is 100~ms) to each communication link using traffic control rules\footnote{https://man7.org/linux/man-pages/man8/tc.8.html} and control groups of the Linux operating system\footnote{https://man7.org/linux/man-pages/man7/cgroups.7.html}.

Fanout is the number of nodes that are selected as gossip targets by a node at each gossip step in order
to retransmit the message~\cite{leitao2010gossip}. The fanout parameter of gossip communication protocol is set to 8 for Algorand and Bitcoin implementations, and 16 for RapidChain because of IDA gossip. 
In addition, we relied on a custom registry service to bootstrap the system: at startup, a node registers itself to the registry service and receives a list of available nodes from the service.
In all experiments, we did not disseminate transactions in the network. Instead, we assumed that nodes have access to pre-initialized transaction pool to populate block payloads. 

\subsection{\algorandplus Performance Evaluation}
\label{algopp_performance}

We conducted two sets of experiments to evaluate the performance improvements of \algorandplus compared to \algorand.
In the first set of experiments described in Section~\ref{subsec:algorandsmall}, we compare the performance of both protocols using 1,000 nodes. 
In the second set of experiments described in Section~\ref{subsec:algorandlarge}, we deployed up to 10,000 nodes to compare the scalability characteristics of both protocols.

\subsubsection{\algorandplus Latency and Throughput}
\label{subsec:algorandsmall}
To evaluate the normal case performance of Algorand and \algorandplus, we deployed 1,000 nodes on ten machines (100 nodes per machine), and we varied the macroblock size from 1~MB to 24~MB and the concurrency level of \algorandplus from 1 to 32. 
For each experiment, elected leaders build blocks of size the expected macroblock size divided by $C_l$. 
The experiments with the concurrency level 1 ($Cl=1$) exhibit \algorand's behavior.


\begin{figure}
\centering
\begin{minipage}[b]{.4\textwidth}
\includegraphics[width=\columnwidth]{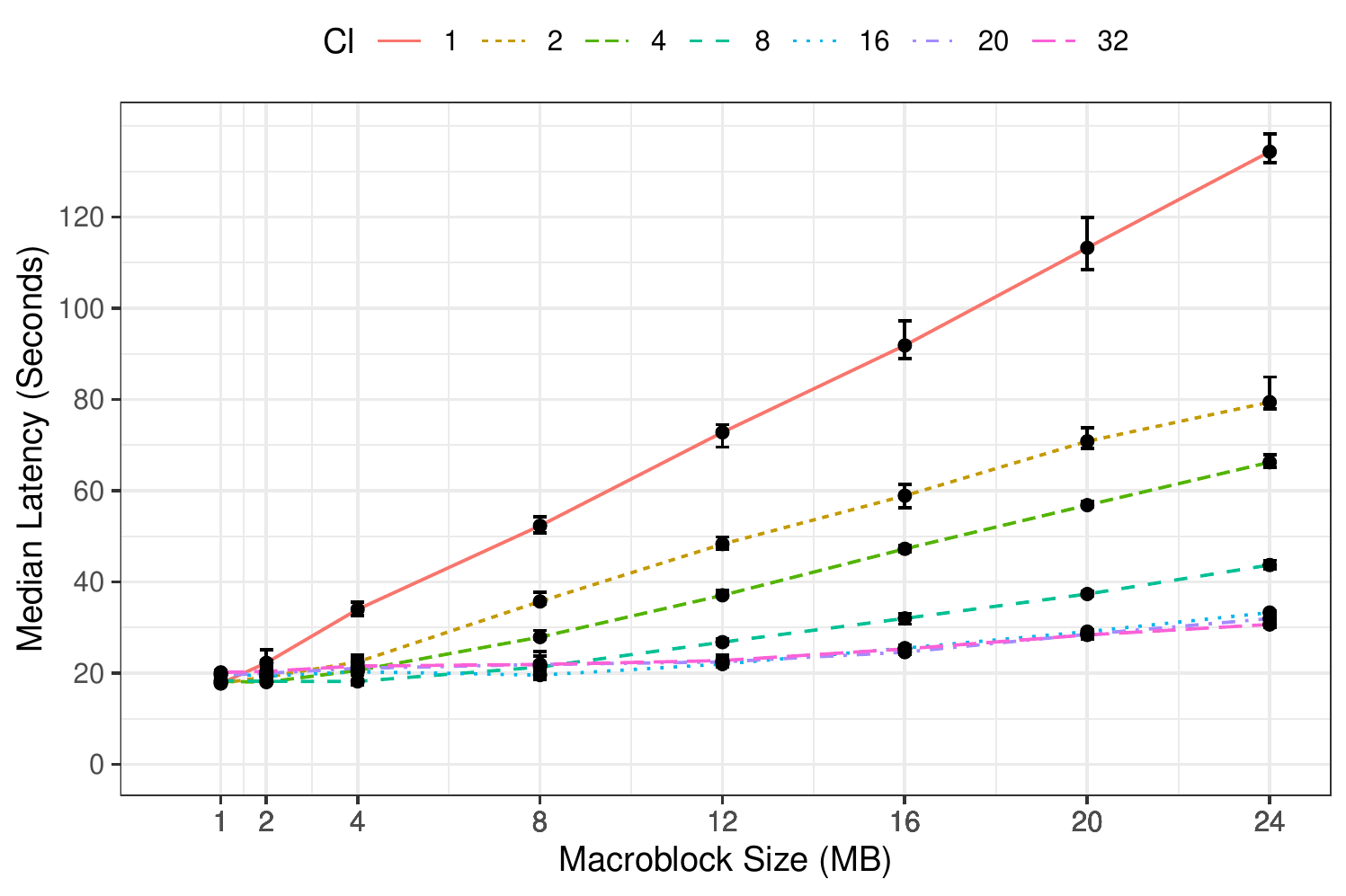}
\end{minipage}\\
\caption{Round latency of \algorandplus with various concurrency levels ($Cl$). $Cl$=1 is Algorand.}\label{fig:round-latency}
\begin{minipage}[b]{.4\textwidth}
\includegraphics[width=\columnwidth]{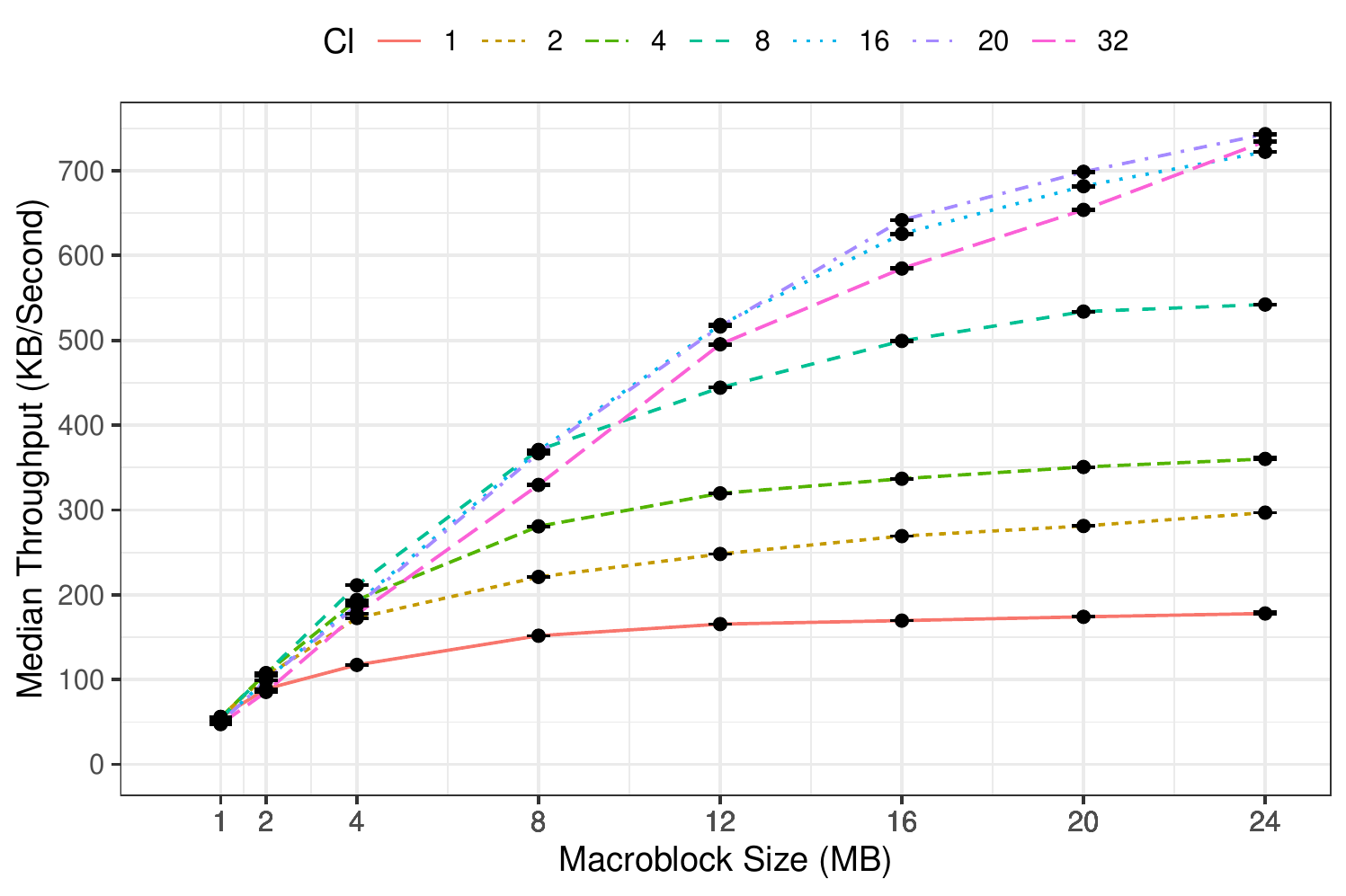}
\end{minipage}
\caption{Effective throughput of \algorandplus with various concurrency levels ($Cl$). $Cl$=1 is Algorand.}\label{fig:effective-throughput}
\vspace{-4mm}
\end{figure}



We start by evaluating the round latency of the two protocols, which indicates how long it takes for a block to be appended to the chain. Round latency is measured as the duration between the proposal time of a block by a leader and the time when all nodes observe this block in the blockchain. Results depicted in Figure~\ref{fig:round-latency} show that the round latency of Algorand increases rapidly as the size of the appended blocks becomes larger. 
Furthermore, we observe that the round latency of \algorandplus largely outperforms the one of Algorand for macroblock sizes larger or equal to 2~MB. 
For instance, for 4~MB blocks, \algorandplus improves the round latency of Algorand from 33.9\% with $Cl$=2 to 46.4\% with $Cl$=8.
The gap between \algorandplus and Algorand becomes even larger with larger block sizes.
Indeed, we observe a round latency improvement varying from 40.9\% with $Cl$=2 to 76.7\% with $Cl$=32 in the configuration with 24~MB blocks.



In addition to round latency, we evaluate the effective throughput of \algorandplus compared to Algorand. 
Since some macroblocks may contain fewer blocks than expected, we cannot derive throughput directly from the round latency and the macroblock size. We define effective throughput as the actual amount of data per second appended to the blockchain. Results of this experiment are depicted in Figure~\ref{fig:effective-throughput}. As shown in this figure, \algorandplus outperforms Algorand in all configurations and can reach up to 743 KB/s. That is four-fold improvement using 24~MB blocks, and $Cl$=20.

Our results also highlight the limits of multiplexing Algorand instances. 
Indeed, we reach the maximum throughput with $Cl=20$, and increasing the concurrency level does not improve performance any further. 

 %

\begin{figure}
\centering
\begin{minipage}[b]{.4\textwidth}
\includegraphics[width=\columnwidth]{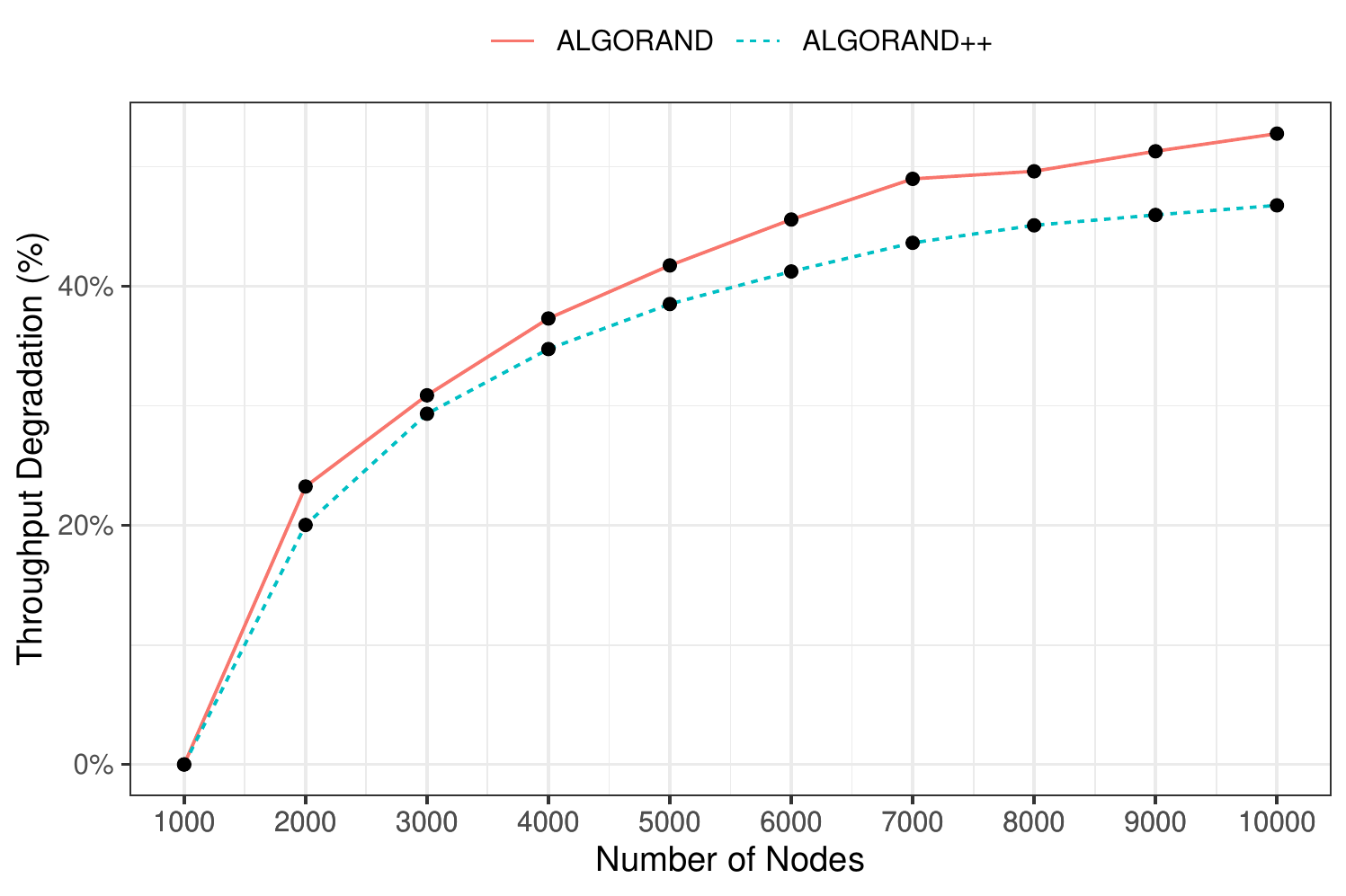}
    \caption{Effective throughput degradation}
    \label{res:scale_thr}
\end{minipage}\\
\begin{minipage}[b]{.4\textwidth}
\includegraphics[width=\columnwidth]{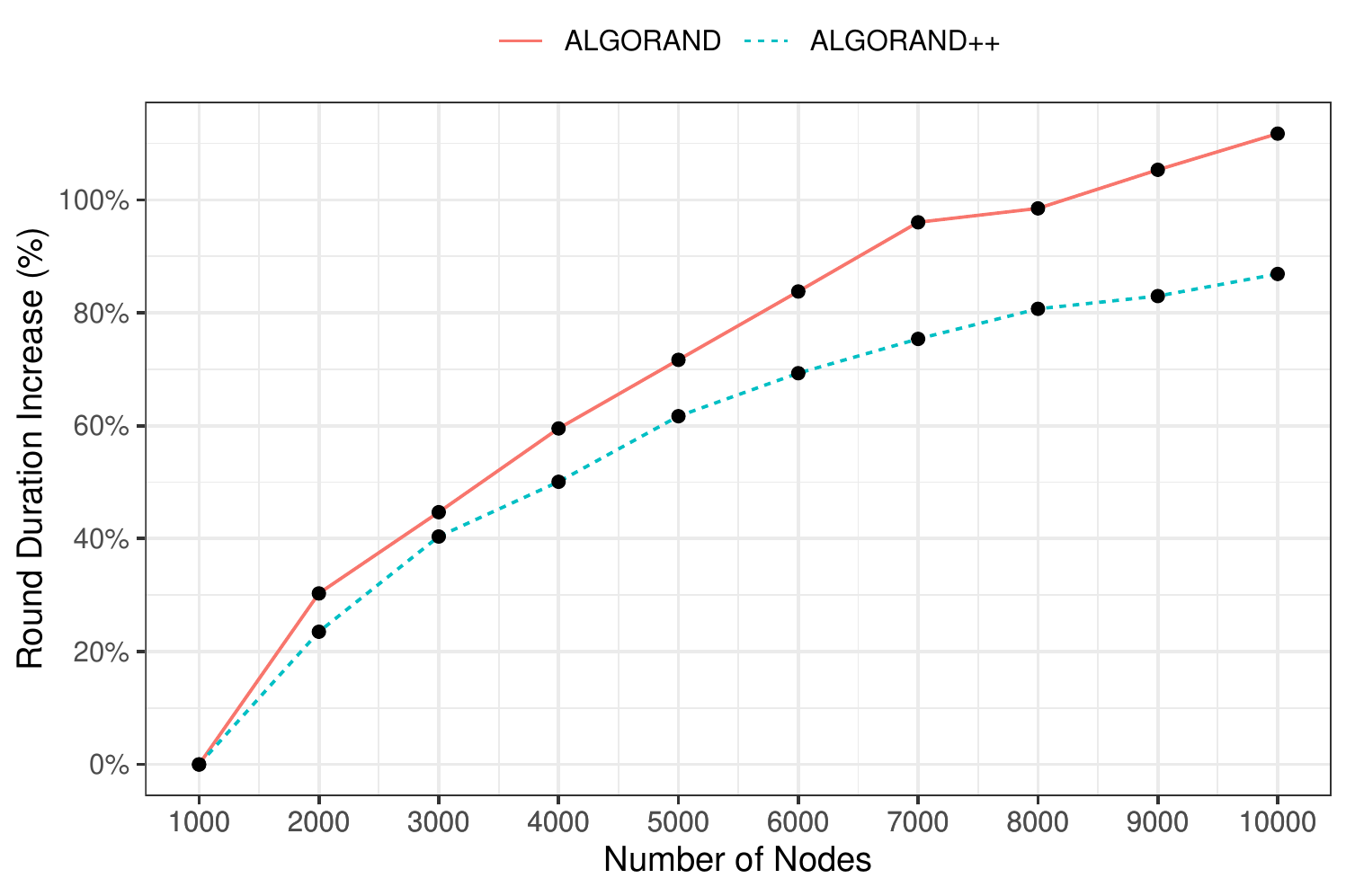}
    \caption{Round duration increase}
    \label{res:scale_lat}
\end{minipage}
\vspace{-4mm}
\end{figure}

\subsubsection{\algorandplus evaluation at scale}
\label{subsec:algorandlarge}

To evaluate how \algorandplus and Algorand scale, we vary the number of machines in the testbed from 10 to 100 with 100 nodes per machine, allowing us to emulate up to 10,000 nodes. In Algorand experiments, we use 1~MB blocks. In \algorandplus experiments, we use 20~MB blocks and $Cl=20$. We then measure the throughput degradation and round latency increase as a function of the number of nodes in the system compared to the baseline with 1,000 nodes. Regarding throughput, results depicted in Figure~\ref{res:scale_thr} show that similarly to Algorand, \algorandplus suffers a throughput degradation when the number of nodes increases in the system. Nevertheless, the mechanisms brought by \protoname to Algorand do not degrade its scalability. It rather improves it when the number of nodes increases as illustrated by the configuration with 10,000 nodes where a 6\% lower degradation difference can be observed for \algorandplus compared to Algorand.

Results related to the latency increase with respect to the number of nodes in the system are depicted in Figure~\ref{res:scale_lat}. From this figure, we observe that the gap in terms of round duration increases between Algorand and \algorandplus with the number of nodes. This illustrates the fact that \algorandplus exhibits better scalability than Algorand. 


\subsection{\rapidchainplus Performance Evaluation}
\label{rapidpp_performance}
We conducted a set of experiments to evaluate the performance of \rapidchainplus compared to \rapidchain in terms of round latency (Section~\ref{rapidchainpp_latency}) and effective throughput (Section~\ref{rapidchainpp_thr}). 
Because the throughput of \rapidchain is the sum of throughput of each shard, we have considered a single shard deployment scenario.
In the original paper, \rapidchain considers a committee size of up to 250 nodes. 
We applied this configuration and deployed 250 nodes on 10 machines. 



\begin{figure}
\centering
\begin{minipage}[b]{.4\textwidth}
   \includegraphics[width=\columnwidth]{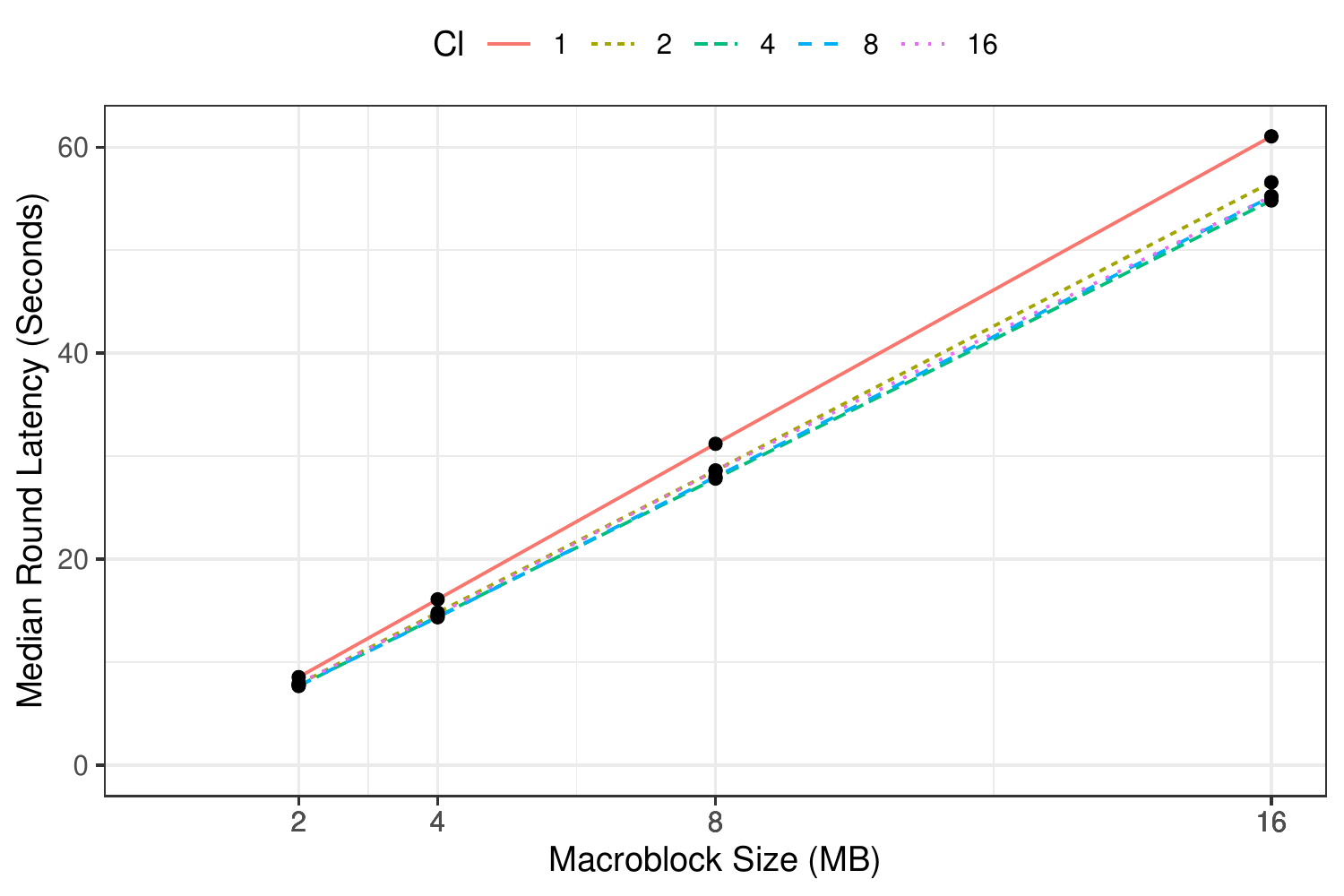}
\end{minipage}\\
\caption{Latency  of \rapidchainplus with various concurrency levels ($Cl$). $Cl$=1 is RapidChain.}
\label{fig:latency-rapidchain}
\begin{minipage}[b]{.4\textwidth}
 \includegraphics[width=\columnwidth]{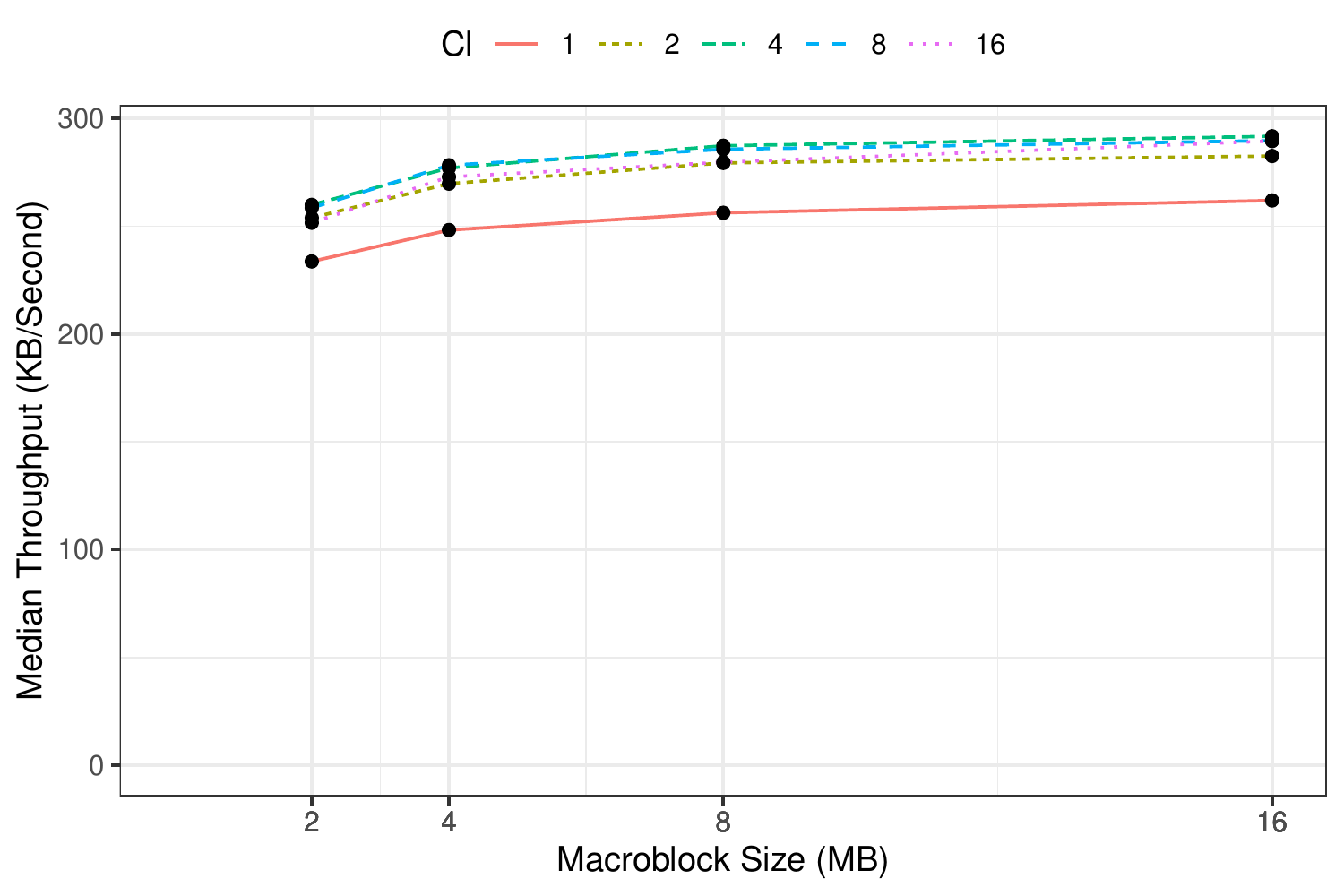}
\end{minipage}
\caption{Throughput of \rapidchainplus with various concurrency levels ($Cl$). $Cl$=1 is RapidChain.}
\label{fig:throughput-rapidchain}
\vspace{-2mm}
\end{figure}

In our experiments,
we vary the macroblock size from 2 to 16 MB, and for each macroblock size, we vary the concurrency level $Cl$ from 1 to 16. 
Similar to \algorand, elected leaders build blocks of size the expected macroblock size divided by $Cl$. 
Each experiment lasted 20 rounds, and we measured the median throughput and latency values as observed by individual nodes. 




\subsubsection{\rapidchainplus round latency}
\label{rapidchainpp_latency}
Figure~\ref{fig:latency-rapidchain} shows the measured round latency for \rapidchainplus with various concurrency levels. 
In the figure, $Cl=1$ corresponds to the latency measurements of the original protocol.
From this figure we observe that for all macroblock sizes, the round latency provided by \rapidchainplus is 10\% lower than the provided round latency of \rapidchain. 
All \rapidchainplus instances provide similar latency values.

\subsubsection{\rapidchainplus effective throughput}
\label{rapidchainpp_thr}
Figure~\ref{fig:throughput-rapidchain} shows the throughput measurements for \rapidchainplus compared to \rapidchain as a function of the macroblock size. 
From this figure, we observe that the highest obtained throughput with \rapidchain is 262 KB/s with the 16-MB block size. 
Instead, \rapidchainplus's highest throughput is 292 KB/s obtained with macroblock size of 16~MB and with a concurrency level $Cl=4$. 
For all block sizes, \rapidchainplus provides a throughput increase of approximately 10\% compared with \rapidchain. 


In our experiments, we can observe that \rapidchainplus provides latency and throughput results 10\% better compared to the ones of \rapidchain.



\subsection{\bitcoinplus Performance Evaluation}
\label{bitcoinpp_performance}
\bitcoinplus reduces the difficulty of cryptographic puzzles inversely proportional to $Cl$ value to elect $Cl$ leaders on average every 10 minutes. 
We conducted experiments to evaluate the performance of \bitcoinplus compared to the ones of \bitcoin with similar cryptographic puzzle difficulty. 
In our experiments we vary the concurrency level of \bitcoinplus from 2 to 96. 
For each concurrency level $Cl$, the associated cryptographic puzzle difficulty is set for both \bitcoin and \bitcoinplus so as to obtain $Cl$ leaders every 10 minutes on average. 
We considered macroblocks of size 1~MB and 500~KB.
Contrarily to the approaches in \rapidchainplus and \algorandplus, we considered smaller macroblock sizes. 
Indeed, keeping high macroblock size would impact the protocol performance because of the time it takes to gossip blocks to the entire system in comparison with the frequency at which nodes find solution to the cryptographic puzzles.
For this reason, we maintain relatively small macroblock size (1 MB and 500 KB).
The size of each block is, as in the \rapidchainplus and \algorandplus, computed as the expected macroblock size divided by the concurrency level $Cl$.




\begin{figure}
\centering
\begin{minipage}[b]{.4\textwidth}
    \includegraphics[width=\columnwidth]{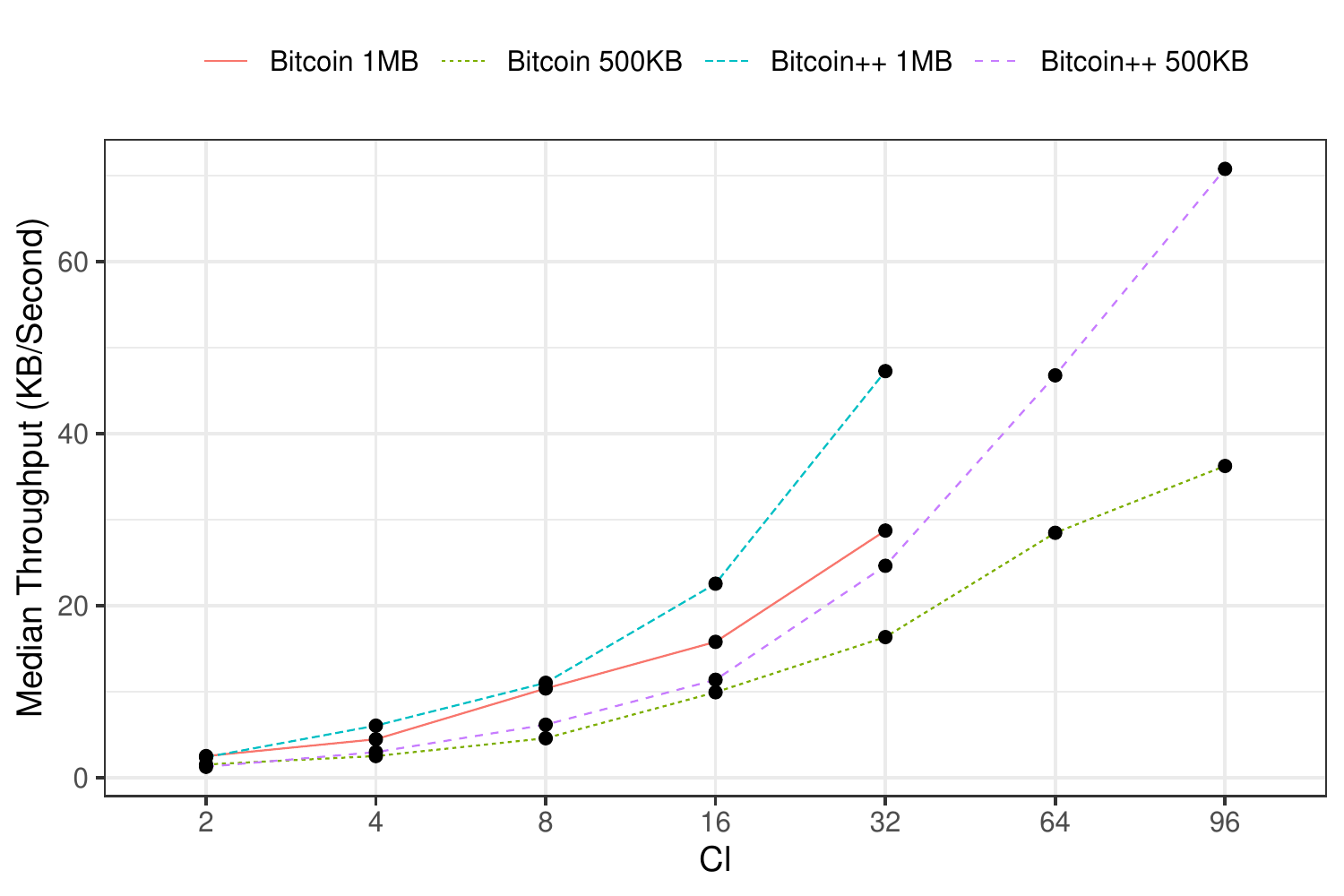}
\end{minipage}\\
\caption{Throughput of Bitcoin Reduced Difficulty and \bitcoinplus. Higher $Cl$ means lower difficulty.}
\label{fig:throughput-bitcoin}
\begin{minipage}[b]{.4\textwidth}
  \includegraphics[width=\columnwidth]{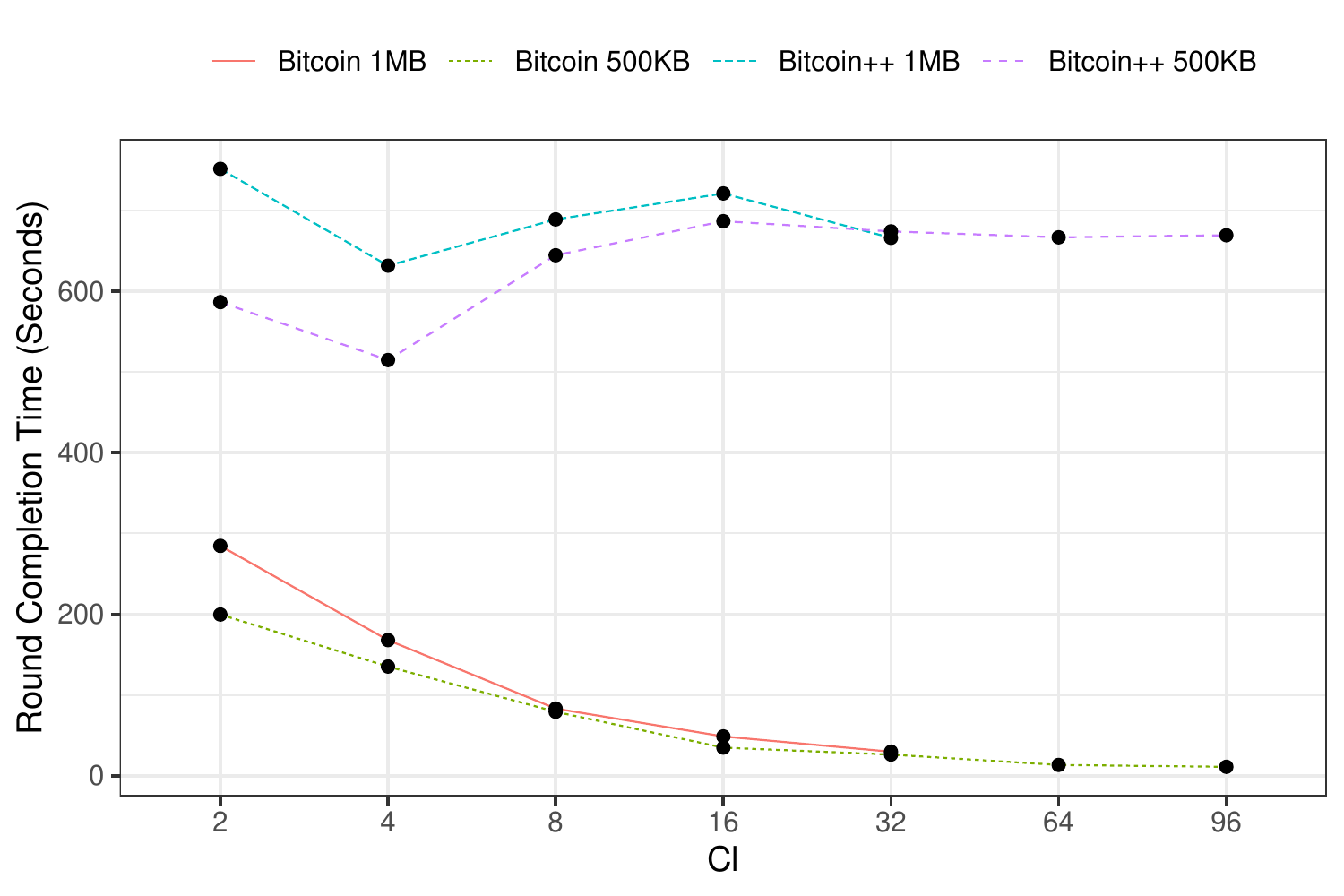}
\end{minipage}
\caption{Latency of Bitcoin Reduced Difficulty and \bitcoinplus. Higher $Cl$ means lower difficulty.}
\label{fig:roundcompletion-bitcoin}
\end{figure}


\subsubsection{\bitcoinplus effective throughput}
Figure~\ref{fig:throughput-bitcoin} shows median throughput figures for both protocols.
The configuration with 1-MB block size and $Cl=1$ that actually represent the original \bitcoin protocol (not represented in the plots) provides a throughput of 1.67KB/s with a latency of 600 seconds).
In all experiments, \bitcoinplus provides superior throughput compared to \bitcoin with reduced difficulty. 
For the 1-MB block size experiments with $Cl=32$, \bitcoinplus provides 47.3 KB/s, a \%65 throughput improvement compared to the associated reduced difficulty \bitcoin with its 28.7 KB/s.
Using 1~MB blocks, we could not increase the $Cl$ value further because of the block dissemination time. 
Regarding the experiments with the 500-KB block size, we observe similar trends.
\bitcoinplus can provide up 70.8 KB/s with $Cl=96$, a two-fold throughput increase compared with \bitcoin and its 35.3 KB/s.


Although decreasing the cryptographic puzzle difficulty seems to help \bitcoin and \bitcoinplus to increase their throughput, it also impacts the number forks, as well as the number of blocks appended to the chain that are then removed when forks are detected. 
To measure such impact, we observe the ratio between the data produced by the leaders and the data actually appended to the blockchain at the end of the experiment.
This measure helps understand the extent of fork occurences in each protocol, which also represents the payload of transactions appended to the chain and later discarded by the nodes.
Our results show that regardless of the concurrency level of \bitcoinplus, \bitcoinplus produces approximately 20\% more data than what it actually appends to the chain.
As for the \bitcoin case, when the cryptographic puzzle difficulty decreases, the latter ratio increases up to 120\%, meaning that \bitcoin produces 1.2 additional blocks of transactions for each block it appends to the chain.

\subsubsection{\bitcoinplus round latency}

Figure~\ref{fig:roundcompletion-bitcoin} shows the median round completion times for \bitcoinplus compared to Bitcoin. 
Because of the choice we made to evaluate the performance of the \bitcoinplus protocol compared to Bitcoin with reduced difficulty, it is natural that round completion time of Bitcoin with reduced difficulty is lower. 
On the contrary, \bitcoinplus appends a single macroblock every 10 minutes, which corresponds to 600 seconds round completion time.

In all experiments, Bitcoin++ provides superior throughput compared to Bitcoin with reduced difficulty. 
Additionally, \bitcoinplus discards significantly fewer data and reduces the occurence of forks in comparison with \bitcoin.



%% file: section/conclusion.tex
\section{Conclusion}
\label{sec:conclusion}

We presented \protoname, a set of reusable principles to multiplex leader-based blockchain consensus protocols in order to improve their performance.
To assess the improvements brought by \protoname on existing blockchain protocols, we applied its principles on three major blockchains:
Algorand, Rapidchain and Bitcoin. 
We presented how \protoname's principles apply to these blockchains and evaluated the performance of the resulting blokchains.
Our evaluation, involving up to 10,000 nodes deployed on 100 physical machines, shows that using \protoname can provide up to a 300\% improvement in both throughput and latency reduction.

%% file: alder-ICDSC-2022.bbl
\begin{thebibliography}{10}
\providecommand{\url}[1]{#1}
\csname url@samestyle\endcsname
\providecommand{\newblock}{\relax}
\providecommand{\bibinfo}[2]{#2}
\providecommand{\BIBentrySTDinterwordspacing}{\spaceskip=0pt\relax}
\providecommand{\BIBentryALTinterwordstretchfactor}{4}
\providecommand{\BIBentryALTinterwordspacing}{\spaceskip=\fontdimen2\font plus
\BIBentryALTinterwordstretchfactor\fontdimen3\font minus
  \fontdimen4\font\relax}
\providecommand{\BIBforeignlanguage}[2]{{%
\expandafter\ifx\csname l@#1\endcsname\relax
\typeout{** WARNING: IEEEtran.bst: No hyphenation pattern has been}%
\typeout{** loaded for the language `#1'. Using the pattern for}%
\typeout{** the default language instead.}%
\else
\language=\csname l@#1\endcsname
\fi
#2}}
\providecommand{\BIBdecl}{\relax}
\BIBdecl

\bibitem{nakamotoBitcoin2008}
S.~Nakamoto, ``Bitcoin: A {{Peer}}-to-{{Peer Electronic Cash System}},'' p.~9,
  2008.

\bibitem{androulakiHyperledger2018}
E.~Androulaki \emph{et~al.}, ``Hyperledger fabric: A distributed operating
  system for permissioned blockchains,'' in \emph{Proceedings of the
  {{Thirteenth EuroSys Conference}} on - {{EuroSys}} '18}.\hskip 1em plus 0.5em
  minus 0.4em\relax {Porto, Portugal}: {ACM Press}, 2018, pp. 1--15.

\bibitem{giladAlgorandScalingByzantine2017}
Y.~Gilad, R.~Hemo, G.~Vlachos, and N.~Zeldovich, ``Algorand: Scaling
  {{Byzantine Agreements}} for {{Cryptocurrencies}},'' in \emph{Proceedings of
  the 26th {{Symposium}} on {{Operating Systems Principles}} - {{SOSP}}
  '17}.\hskip 1em plus 0.5em minus 0.4em\relax {Shanghai, China}: {ACM Press},
  2017, pp. 51--68.

\bibitem{noauthor_hyperledger_nodate}
\BIBentryALTinterwordspacing
``Hyperledger – {Open} {Source} {Blockchain} {Technologies}.'' [Online].
  Available: \url{https://www.hyperledger.org/}
\BIBentrySTDinterwordspacing

\bibitem{uriarteBlockchainBased2018}
R.~B. Uriarte and R.~DeNicola, ``Blockchain-{{Based Decentralized Cloud}}/{{Fog
  Solutions}}: Challenges, {{Opportunities}}, and {{Standards}},'' \emph{IEEE
  Communications Standards Magazine}, vol.~2, no.~3, pp. 22--28, Sep. 2018.

\bibitem{raval2016decentralized}
S.~Raval, \emph{Decentralized applications: harnessing Bitcoin's blockchain
  technology}.\hskip 1em plus 0.5em minus 0.4em\relax " O'Reilly Media, Inc.",
  2016.

\bibitem{belchior2022survey}
\BIBentryALTinterwordspacing
R.~Belchior, A.~Vasconcelos, S.~Guerreiro, and M.~Correia, ``A survey on
  blockchain interoperability: Past, present, and future trends,'' vol.~54,
  no.~8, 2021. [Online]. Available: \url{https://doi.org/10.1145/3471140}
\BIBentrySTDinterwordspacing

\bibitem{guerraoui2019consensus}
R.~Guerraoui \emph{et~al.}, ``The consensus number of a cryptocurrency,'' in
  \emph{Proceedings of the 2019 ACM Symposium on Principles of Distributed
  Computing}, 2019, pp. 307--316.

\bibitem{wood2014ethereum}
G.~Wood \emph{et~al.}, ``Ethereum: A secure decentralised generalised
  transaction ledger,'' \emph{Ethereum project yellow paper}, vol. 151, no.
  2014, pp. 1--32, 2014.

\bibitem{eyalBitcoinNG2016}
I.~Eyal, A.~E. Gencer, E.~G. Sirer, and R.~{van Renesse}, ``Bitcoin-{{NG}}: A
  {{Scalable Blockchain Protocol}},'' \emph{13th USENIX Symposium on Networked
  Systems Design and Implementation (NSDI '16)}, p.~16, 2016.

\bibitem{kogias2016enhancing}
E.~K. Kogias \emph{et~al.}, ``Enhancing bitcoin security and performance with
  strong consistency via collective signing,'' in \emph{25th $\{$usenix$\}$
  security symposium ($\{$usenix$\}$ security 16)}, 2016, pp. 279--296.

\bibitem{zamaniRapidChain2018}
M.~Zamani \emph{et~al.}, ``{{RapidChain}}: Scaling {{Blockchain}} via {{Full
  Sharding}},'' in \emph{Proceedings of the 2018 {{ACM SIGSAC Conference}} on
  {{Computer}} and {{Communications Security}}}.\hskip 1em plus 0.5em minus
  0.4em\relax {Toronto Canada}: {ACM}, Oct. 2018, pp. 931--948.

\bibitem{yuOHIE2020}
H.~Yu, I.~Nikoli{\'c}, R.~Hou, and P.~Saxena, ``{{OHIE}}: Blockchain {{Scaling
  Made Simple}},'' in \emph{2020 {{IEEE Symposium}} on {{Security}} and
  {{Privacy}} ({{SP}})}, May 2020, pp. 90--105.

\bibitem{abraham2016solida}
I.~Abraham, D.~Malkhi, K.~Nayak, L.~Ren, and A.~Spiegelman, ``Solida: A
  blockchain protocol based on reconfigurable byzantine consensus,''
  \emph{arXiv preprint arXiv:1612.02916}, 2016.

\bibitem{davidOuroboros2018}
B.~David \emph{et~al.}, ``Ouroboros {{Praos}}: An {{Adaptively}}-{{Secure}},
  {{Semi}}-synchronous {{Proof}}-of-{{Stake Blockchain}},'' in \emph{Advances
  in {{Cryptology}} \textendash{} {{EUROCRYPT}} 2018}, ser. Lecture {{Notes}}
  in {{Computer Science}}, J.~B. Nielsen and V.~Rijmen, Eds.\hskip 1em plus
  0.5em minus 0.4em\relax {Cham}: {Springer International Publishing}, 2018,
  pp. 66--98.

\bibitem{eyal2016bitcoin}
I.~Eyal, A.~E. Gencer, E.~G. Sirer, and R.~Van~Renesse, ``Bitcoin-ng: A
  scalable blockchain protocol,'' in \emph{13th $\{$USENIX$\}$ symposium on
  networked systems design and implementation ($\{$NSDI$\}$ 16)}, 2016, pp.
  45--59.

\bibitem{syta2016keeping}
E.~Syta \emph{et~al.}, ``Keeping authorities" honest or bust" with
  decentralized witness cosigning,'' in \emph{2016 IEEE Symposium on Security
  and Privacy (SP)}.\hskip 1em plus 0.5em minus 0.4em\relax Ieee, 2016, pp.
  526--545.

\bibitem{zamani2018rapidchain}
M.~Zamani \emph{et~al.}, ``Rapidchain: Scaling blockchain via full sharding,''
  in \emph{Proceedings of the 2018 ACM SIGSAC Conference on Computer and
  Communications Security}, 2018, pp. 931--948.

\bibitem{wang2019monoxide}
J.~Wang and H.~Wang, ``Monoxide: Scale out blockchains with asynchronous
  consensus zones,'' in \emph{16th $\{$USENIX$\}$ Symposium on Networked
  Systems Design and Implementation ($\{$NSDI$\}$ 19)}, 2019, pp. 95--112.

\bibitem{elasticoluu2016}
L.~Luu \emph{et~al.}, ``A {{Secure Sharding Protocol For Open Blockchains}},''
  in \emph{Proceedings of the 2016 {{ACM SIGSAC Conference}} on {{Computer}}
  and {{Communications Security}}}, ser. {{CCS}} '16.\hskip 1em plus 0.5em
  minus 0.4em\relax {New York, NY, USA}: {Association for Computing Machinery},
  Oct. 2016, pp. 17--30.

\bibitem{kokoris2018omniledger}
E.~Kokoris-Kogias \emph{et~al.}, ``Omniledger: A secure, scale-out,
  decentralized ledger via sharding,'' in \emph{2018 IEEE Symposium on Security
  and Privacy (SP)}.\hskip 1em plus 0.5em minus 0.4em\relax IEEE, 2018, pp.
  583--598.

\bibitem{lamport2001paxos}
L.~Lamport \emph{et~al.}, ``Paxos made simple,'' \emph{ACM Sigact News},
  vol.~32, no.~4, pp. 18--25, 2001.

\bibitem{barcelona2008mencius}
C.-S. Barcelona, ``Mencius: building efficient replicated state machines for
  wans,'' in \emph{8th USENIX Symposium on Operating Systems Design and
  Implementation (OSDI 08)}, 2008.

\bibitem{milosevic2013bounded}
Z.~Milosevic, M.~Biely, and A.~Schiper, ``Bounded delay in byzantine-tolerant
  state machine replication,'' in \emph{2013 IEEE 32nd International Symposium
  on Reliable Distributed Systems}.\hskip 1em plus 0.5em minus 0.4em\relax
  IEEE, 2013, pp. 61--70.

\bibitem{castro1999practical}
M.~Castro, B.~Liskov \emph{et~al.}, ``Practical byzantine fault tolerance,'' in
  \emph{OSDI}, vol.~99, no. 1999, 1999, pp. 173--186.

\bibitem{stathakopoulou2019mir}
C.~Stathakopoulou, T.~David, and M.~Vukolic, ``Mir-bft: High-throughput bft for
  blockchains,'' \emph{arXiv preprint arXiv:1906.05552}, 2019.

\bibitem{yu2020ohie}
H.~Yu, I.~Nikoli{\'c}, R.~Hou, and P.~Saxena, ``Ohie: Blockchain scaling made
  simple,'' in \emph{2020 IEEE Symposium on Security and Privacy (SP)}.\hskip
  1em plus 0.5em minus 0.4em\relax IEEE, 2020, pp. 90--105.

\bibitem{crain2021red}
T.~Crain, C.~Natoli, and V.~Gramoli, ``Red belly: a secure, fair and scalable
  open blockchain,'' in \emph{Proceedings of the 42nd IEEE Symposium on
  Security and Privacy (S\&P’21)}, 2021.

\bibitem{antoniadis2021leaderless}
K.~Antoniadis, A.~Desjardins, V.~Gramoli, R.~Guerraoui, and M.~I. Zablotchi,
  ``Leaderless consensus,'' in \emph{IEEE 41st International Conference on
  Distributed Computing Systems(ICDCS)}, 2021.

\bibitem{rocket2019scalable}
T.~Rocket, M.~Yin, K.~Sekniqi, R.~van Renesse, and E.~G. Sirer, ``Scalable and
  probabilistic leaderless bft consensus through metastability,'' \emph{arXiv
  preprint arXiv:1906.08936}, 2019.

\bibitem{micaliVerifiableRandomFunctions1999}
S.~Micali, M.~Rabin, and S.~Vadhan, ``Verifiable random functions,'' in
  \emph{40th {{Annual Symposium}} on {{Foundations}} of {{Computer Science}}
  ({{Cat}}. {{No}}.{{99CB37039}})}, Oct. 1999, pp. 120--130.

\bibitem{maymounkov2002kademlia}
P.~Maymounkov and D.~Mazieres, ``Kademlia: A peer-to-peer information system
  based on the xor metric,'' in \emph{International Workshop on Peer-to-Peer
  Systems}.\hskip 1em plus 0.5em minus 0.4em\relax Springer, 2002, pp. 53--65.

\bibitem{ren2017practical}
L.~Ren, K.~Nayak, I.~Abraham, and S.~Devadas, ``Practical synchronous byzantine
  consensus,'' \emph{arXiv preprint arXiv:1704.02397}, 2017.

\bibitem{bolze2006grid}
R.~Bolze \emph{et~al.}, ``Grid'5000: a large scale and highly reconfigurable
  experimental grid testbed,'' \emph{The International Journal of High
  Performance Computing Applications}, vol.~20, no.~4, pp. 481--494, 2006.

\bibitem{leitao2010gossip}
J.~Leitao, J.~Pereira, and L.~Rodrigues, ``Gossip-based broadcast,'' in
  \emph{Handbook of Peer-to-Peer Networking}.\hskip 1em plus 0.5em minus
  0.4em\relax Springer, 2010, pp. 831--860.

\end{thebibliography}
